\documentclass[12pt]{article}
\pdfoutput=1
\usepackage{jheppub}
\usepackage[usenames,dvipsnames]{xcolor}
\usepackage{colortbl}
\usepackage{tikz}
\usepackage{enumitem}
\usetikzlibrary{matrix,arrows,positioning,shapes,snakes,fadings,decorations.pathmorphing,
decorations.pathreplacing,automata,shadows,decorations.markings,patterns,decorations.text}
\usepackage{bm,upgreek}
\usepackage{pgfplots}

\usepackage{amsmath,amssymb,euscript,array,mathrsfs,appendix,ctable,mathabx,amsthm}
\usepackage{mathtools,float}

\usepackage{graphicx}
\usepackage[small]{caption}

\usepackage{titlesec}
\titleformat{\subsection}[display]{\it}{}{0.1cm}{\vspace{-1.5cm}\begin{center}\thesubsection\hspace{0.2cm}}[\end{center}\vspace{-0.5cm}]

\newcommand\arXiv[1]{\href{http://arxiv.org/abs/#1}{\tt arXiv:#1}} 

\def\ms{\text{sh}\tfrac\pi\beta}

\newcommand*\circled[1]{\tikz[baseline=(char.base)]{%
            \node[shape=circle,fill=blue!10,draw,inner sep=1pt] (char) {#1};}}

\newcommand{\EQ}[1]{\begin{equation}\begin{split} #1
\end{split}\end{equation}}
\def\ben
{\begin{equation}}
\def\een{\end{equation}}

    \let\L=\Lambda
   
 \let\W=\mu

\def\W={\cal W}
\def\L ={\cal L}

\def\be{\begin{equation}}
\def\ee{\end{equation}}
\def\ba{\begin{array}}
\def\ea{\end{array}}

\def\dalemb#1#2{{\vbox{\hrule height .#2pt
        \hbox{\vrule width.#2pt height#1pt \kern#1pt
                \vrule width.#2pt}
        \hrule height.#2pt}}}

\newcommand{\bea}{\begin{eqnarray}}
\newcommand{\eea}{\end{eqnarray}}

\newcommand{\ext}{\text{ext}}

\title{Ephemeral Islands, Plunging Quantum Extremal Surfaces and BCFT channels}
\author{Timothy J. Hollowood, S. Prem Kumar, Andrea Legramandi and Neil Talwar}
\affiliation{Department of Physics,\\
Swansea University,\\
Singleton Park, Swansea,\\
SA2 8PP, U.K.}
\emailAdd{t.j.hollowood@swansea.ac.uk, s.p.kumar@swansea.ac.uk, andrea.legramandi@swansea.ac.uk,n.talwar.2017429@swansea.ac.uk}
\abstract{We consider entanglement entropies of finite spatial intervals in Minkowski radiation baths coupled to the eternal black hole in JT gravity, and the related problem involving free fermion BCFT in the thermofield double state. We show that the non-monotonic entropy evolution in the black hole problem precisely matches that of the free fermion theory in a high temperature limit, and the results have the form   expected for CFTs with quasiparticle description. Both exhibit rich behaviour that  involves at intermediate times, an entropy saddle with an island in the former case, and in the latter a special class of disconnected OPE channels. The quantum extremal surfaces start inside the horizon, but can emerge from and plunge back inside as time evolves, accompanied by a characteristic dip in the entropy also seen in the free fermion BCFT. Finally an entropy equilibrium is reached with a no-island saddle. %In the limit that the results match those of the free fermion BCFT, the entanglement wedge reconstruction of the interior of the black hole becomes simple.
}

\begin{document}
\maketitle

\newpage

\section{Introduction}

Information recovery from  Hawking radiation emitted by an evaporating black hole has long presented a  fascinating puzzle \cite{Hawking:1974sw,Hawking:1976ra}. Its clearest quantitative formulation is in terms of the von Neumann entropy of the emitted radiation which is expected to follow the Page curve \cite{Page:1993wv,Page:2013dx}, and the information puzzle then translates into the challenge to produce the Page curve  from first principles. Recent progress in obtaining the Page curve for black holes within semiclassical gravity \cite{Penington:2019kki, Almheiri:2019qdq,Almheiri:2020cfm}  revolves around the appearance of new replica wormhole saddle points for  von Neumann entropies, responsible for turning around the Page curve for Hawking radiation as expected from unitarity. The physics of the new saddles is encapsulated within a generalized or {\em fine grained}  entropy formula \cite{Engelhardt:2014gca}, anticipated from holographic considerations \cite{Ryu:2006bv,Hubeny:2007xt,Faulkner:2013ana,Engelhardt:2014gca}. The fine grained  entropy restores the subtle correlations, which were absent in Hawking's original calculation, between the Hawking modes from an old black hole and its early radiation. The underlying mechanism hinges on appending to the semiclassical Hawking radiation $R$, island regions $I$ extending behind the horizon and bounded  by Quantum Extremal Surfaces (QES). In particular, the island is contained within the entanglement wedge of the radiation $R$.\footnote{Various aspects of islands and their appearance in a variety of gravitating setups have been explored in \cite{Almheiri:2019hni,Almheiri:2019yqk,Almheiri:2019psy,Geng:2021iyq, Bhattacharya:2021jrn,Kawabata:2021hac,Bousso:2021sji,Wang:2021woy,Karananas:2020fwx,Hayden:2020vyo,Basak:2020aaa,Choudhury:2020hil,Colin-Ellerin:2020mva,Goto:2020wnk,Matsuo:2020ypv,Hernandez:2020nem, Bhattacharya:2020uun, Ling:2020laa,Chen:2020hmv, Johnson:2020mwi,Hollowood:2020kvk,Chen:2020jvn,Chandrasekaran:2020qtn,Li:2020ceg,Chen:2020uac,Alishahiha:2020qza,Hashimoto:2020cas,Giddings:2020yes,Anegawa:2020ezn,Gautason:2020tmk,Chen:2020wiq,Bhattacharya:2020ymw,Chen:2019iro, Krishnan:2020oun,Ghosh:2021axl}.}

This leads us to one of the central questions in this picture: how precisely are the islands encoded in $R$ and how do the islands implement the `accounting trick' needed to get entropies right?  To understand this we need to know how the semiclassical island prescription ties in with a microscopic description of the evaporating black hole. 
In this work, we will employ a tractable microscopic description of  evaporating black holes in a certain limit, to identify how the nontrivial time evolution of entanglement in the microscopic theory is precisely captured  by an emergent island saddle prescription.  

Such a  microscopic framework is readily available in the useful, simplifying scenario when restricting to the $s$-wave sector of the near horizon geometry of a near-extremal charged black hole.
The evaporating black hole can be described by an effective two dimensional model \cite{Almheiri:2014cka} namely, Jackiw-Teitelboim (JT) gravity \cite{Jackiw:1984je,Teitelboim:1983ux} on AdS$_2$ coupled to  radiation degrees of freedom propagating in AdS$_2$ as well as an asymptotic Minkowski bath region \cite{Almheiri:2019qdq, Almheiri:2019yqk, Almheiri:2019hni}. In this model the bath, represented by massless fields or a conformal field theory (CFT) in 1+1 dimensional Minkowski spacetime, is nongravitating and glued on to the regularized boundary of the AdS$_2$ region with transparent boundary conditions\footnote{The consistency of the islands picture and the Page curve has been questioned in \cite{Geng:2020fxl, Raju:2020smc,Geng:2021hlu,Chowdhury:2021nxw} for long range theories of gravity i.e. when the bath is gravitating. In the case where AdS gravity is coupled to a non-gravitating bath the graviton is not massless \cite{Geng:2020qvw} and the Page curve for the fine-grained entropy is crisply defined. This does not however mean that this Page curve is irrelevant  in asymptotically flat space when gravity in the bath is dynamical \cite{Krishnan:2020oun}. In particular, it is the answer to questions involving ``not-so-fine-grained" entropies \cite{Ghosh:2021axl}.}.  

The microscopic description of the setup arises via the holographic dual viewpont, as a CFT on the half-line coupled to quantum mechanical degrees of freedom on the boundary \cite{Almheiri:2019hni, Almheiri:2019yqk}.  In an appropriate infrared limit, boundary conformal field theory (BCFT) \cite{Cardy:1984bb, Cardy:2004hm} offers a powerful framework to examine this scenario and the emergence  of a Page curve therein \cite{Sully:2020pza,Geng:2021iyq}.  

\begin{figure}[h]
\begin{center}
\begin{tikzpicture} [scale=0.5]
\draw[dotted] (3,0) -- (3,11);
\draw[dotted] (4.5,0) -- (4.5,11);
\draw[dotted] (7.5,0) -- (7.5,11);
\draw[dotted] (12,0) -- (12,11);
\draw[very thick,black!30] (3.5,11) -- (0,7.5) -- (7.5,7.5) -- (11,11);
\draw[very thick,black!30] (0,4.5) -- (6.5,11);
\draw[very thick,blue!50] (0,0) -- (5.5,11);
\draw[very thick,blue!50] (0,3) -- (8,11);
\draw[very thick,blue!50] (7.5,4.5) -- (1,11);
\draw[very thick,blue!50] (7.5,4.5) -- (14,11);
\draw[very thick,blue!50] (0,9) -- (15,9);
\draw[thick,black!30] (0,0) -- (0,11) -- (15,11) -- (15,0) --cycle;
\draw[very thick] (0,0) -- (3,6) -- (4.5,7.5) -- (7.5,4.5) -- (12,9) -- (15,9);
\draw[red,very thick] (3,6) -- (4.5,7.5) -- (7.5,4.5) -- (12,9);
\node at (1.5,0.9) {\footnotesize(I)};
\node at (3.75,0.9) {\footnotesize(II)};
\node at (6,0.9) {\footnotesize(III)};
\node at (9.75,0.9) {\footnotesize(IV)};
\node at (13.5,0.9) {\footnotesize(V)};
\node at (0,-0.7) {\footnotesize0};
\node at (3,-0.7) {\footnotesize$a$};
\node at (4.5,-0.7) {\footnotesize$\tfrac{b-a}2$};
\node at (7.5,-0.7) {\footnotesize$\tfrac{a+b}2$};
\node at (12,-0.7) {\footnotesize$b$};
\node at (7.25,-1.7) {\footnotesize $t$};
\node [rotate=90] at (-1.3,5.5) {\footnotesize $S(A_L\cup A_R)$};
\node at (11,9.5) {\footnotesize (1)};
\node at (2.3,3) {\footnotesize (2)};
\node at (0.6,4.3) {\footnotesize (3)};
\node at (8.5,6.3) {\footnotesize (4)};
\end{tikzpicture}
\caption{\footnotesize BCFT channels for the von Neumann entropy at high temperature, of which the four in blue (labelled numbers as in figure \ref{fig4}) actually compete for the configuration with $A_L=A_R=[a,b]$ and $3a<b$ ($a=3$, $b=12$ and $\beta=0.1$ shown). The channels in grey are always subdominant. The red portion of the curve, channels (3) and (4), corresponds to the island saddle of the black hole in JT gravity. Temporal regimes (I)-(V) are used in the analysis. The exact result for the free fermion theory lies on top of the black and red curve.}
\label{fig6}
\end{center}
\end{figure}
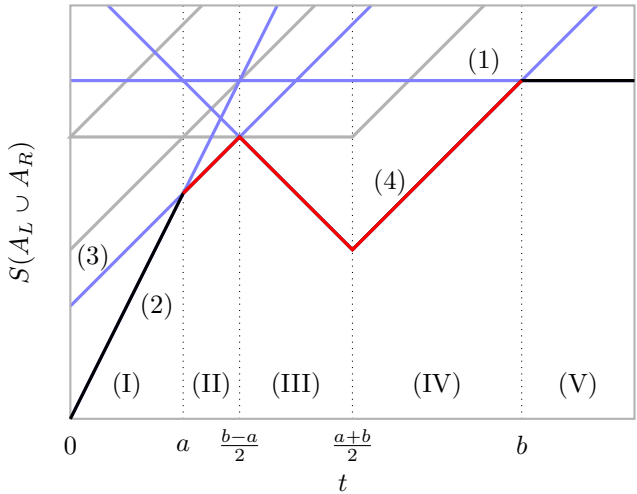

We consider  BCFT prepared in a particularly simple nonequilibrium state, the so-called Thermo-Field Double (TFD) state obtained as a purification of the thermal state. The semiclassical dual description is in terms of the two-sided AdS$_2$ eternal black hole in the Hartle-Hawking state \cite{Israel:1976ur}, coupled to (non-gravitating) radiation baths. The TFD state, although in local thermodynamic equilibrium, is not  in entanglement equilibrium. In this work, we use the two complementary descriptions above to demonstrate:
\begin{enumerate}[label=\protect\circled{{\bf \arabic*}}]
\item{In the high temperature limit, measures of entanglement in BCFT exhibit nontrivial but universal\footnote{By `universal' here, we mean it applies to CFTs with an effective quasiparticle description as explained in \cite{Asplund:2015eha}. This includes free CFTs, rational CFTs and in general those where the high energy density of states is dominated by conserved currents or holomorphic and antiholomorphic operators. } time evolution across several distinctly identifiable temporal  regimes. 
%The exact time evolution of  free fermion BCFT entanglement entropies at high temperatures is reproduced by  competition between distinct OPE (operator product expansion) channels of the relevant correlation functions.
}
\item{The same behaviour emerges via a nontrivial competition between no-island and island saddles in the semiclassical JT gravity description. Island saddles can be `ephemeral' i.e. dominant for short intermediate times and disappear at late time.}
\item{Distinct temporal regimes in entanglement evolution can each be associated to a particular OPE  limit of BCFT correlation functions. Disconnected  channels in BCFT get identified with  corresponding bulk OPE channels (in JT gravity) which involve quantum extremal surfaces that may lie inside, but can then dynamically exit and plunge back inside the horizon during the evolution. }
\end{enumerate}

The situation we focus attention on is the case of two disjoint {\em finite} intervals $A_L$ and $A_R$, 
in the  $L$ and $R$ copies respectively, of the Minkowski bath in the TFD state. The time evolution of the von Neumann entropy is shown in figure \ref{fig6} for the specific case where $A_L$ and $A_R$ are taken to be identical. The von Neumann entropy first increases linearly as modes enter (exit) the  interval in one copy of the bath and their respective purifiers in the TFD state simultaneously exit (enter) the interval in the second bath. The growth slows and experiences a characteristic dip which is controlled by an evolving island saddle in gravity, and distinct (partially) disconnected OPE channels in BCFT. The physics of the dip is simple and follows from a general free quasiparticle picture \cite{Calabrese:2005in, Calabrese:2009qy}. In BCFT, it is due to modes  reflecting off the boundary in either copy, entering $A_L\cup A_R$ along with their respective purifiers in the TFD copy. In the semiclassical gravity description, the dip in  entropy  appears when modes from the bath enter $A_L\cup A_R$ and their purifiers in the TFD are simultaneously captured by the  island behind the horizon. As illustrated in figure \ref{fig17}, the way the entropy is accounted for appears to be different in the BCFT and semiclassical gravity descriptions, but net effect is identical.

A quantitative match between gravity and BCFT occurs when characteristic length and time scales $\sim L$ are taken to be large compared to both the black hole scrambling time scale, and the inverse temperature $\beta$. In terms of the JT gravity parameter $k$ which must be small in the semiclassical limit, we require
\be
L\gg \frac{\beta}{2\pi}\log\frac{2\pi}{\beta k}\gg \beta\ .
\ee
The middle scale here is the scrambling time of the black hole. In this limit the black hole behaves like a mirror and information recovery proceeds by reflection, exactly as in the BCFT, and the boundary entropy $s_b=\log g_b$ \cite{Friedan:2003yc} in the BCFT is identified with the Bekenstein-Hawking entropy of the black hole. If we further impose the requirement $L\gg 1/k$, the system matches the high temperature free fermion theory which has $g_b=1$ (vanishing boundary entropy). In this simple limit, we expect that the black hole interior and quantum extremal surfaces can be reconstructed simply from BCFT data.

 Even away from the strict BCFT limit above, the entanglement evolution in gravity displays the same qualitative and quantitative physics. This indicates that BCFT rules should also apply at high temperatures to the more realistic microscopic description with fully dynamical boundary degrees of freedom (the SYK model) coupled to massless fields (see e.g. \cite{Chen:2020wiq}). 
 
 While this work was in preparation, the preprint \cite{Balasubramanian:2021xcm} appeared which has some overlap with the questions addressed in this paper. In particular, \cite{Balasubramanian:2021xcm} studied the evolution of the entanglement entropy in a holographic large-$c$ BCFT. In this setup non-monotonicity or ``dips" in entanglement entropy are absent since holographic CFTs scramble maximally\footnote{We would like to thank the anonymous referee for drawing our attention to this important physical point.} as seen e.g. in \cite{Asplund:2013zba, Balasubramanian:2011at, Allais:2011ys, Leichenauer:2015xra,Asplund:2015eha}.

The paper is organized as follows. In section \ref{s2} we review basic aspects of the setup of the TFD state and its gravitational analogue, and fix our notation and coordinate systems. Section \ref{sec:bcft} contains the exact analysis of von Neumann entropies in free fermion BCFT and its precise agreement with the general picture of competing OPE channels at high temperatures. We turn to the JT gravity picture in section \ref{sec:gravity} and find the saddle points of the generalized entropy function, and show the precise correspondence between dominant BCFT channels and the pattern of bulk operator product channels involving QES in each temporal regime. In section \ref{s5}, we show that the  high temperature time evolution of entanglement entropies in the BCFT limit can be understood in a simple way in a geometric optics approximation. We conclude with a detailed discussion of the general implications of our findings in section \ref{s6}.

%{\tt Alternative title: Emphemeral Islands with Quantum Extremal Surfaces on the Move}
%{\tt Perhaps summarize the case with infinite intervals as this will already show how BCFT related to JT in an appropriate limit.}

\section{Preliminaries}\label{s2}

We  consider  BCFT on the half-line and take two copies CFT$_L$ and CFT$_R$ prepared in the thermofield double state. The respective spatial coordinates $x$ are positive, and we are interested in spatial intervals of the type $A=A_L\cup A_R$ with $A_L=[a_1,b_1]$, $A_R=[a_2, b_2]$. For simplicity, the initial focus will be on the special case with symmetric intervals $a_1=a_2\equiv a$ and $b_1=b_2\equiv b$.  
%Since JT gravity on AdS$_2$ is holographically dual to a quantum mechanical model, the setup can also be viewed as the CFT on the half-line coupled to quantum mechanical degrees of freedom on the boundary. %In a particular infrared limit, this system can be viewed as boundary conformal field  (BCFT) theory.
%The holographic viewpoint is powerful since, at least in principle, the nonequilibrium evolution of the entanglement structure of the evaporating black hole and its Hawking radiation can be studied within the CFT coupled to the boundary quantum mechanics. In practice, this poses technical challenges, but there is a special simplifying limit, namelu   

%Hawking's information loss paradox is crisply formulated within this setting in terms of an expected Page curve for the von Neumann entropy of the radiation received in the bath regions.

%{\tt Decide on one set of conventions for both problems}

%{\tt Shall we define $y^\pm\to x^\pm$ instead (probably nicer since we don't use Poincar\'e coordinates)?}

\begin{figure}[ht]
\begin{center}
\begin{tikzpicture}[scale=1]
\draw[fill=yellow!20,yellow!20] (-6,0) -- (-3,-3) -- (-3,3) -- cycle;
\draw[fill=yellow!20,yellow!20] (6,0) -- (3,-3) -- (3,3) -- cycle;
\draw[white,fill=Plum!10!white] (-3,-3) -- (3,-3) -- (3,3) -- (-3,3) -- cycle;
\draw[-] (-3,3) -- (-6,0) -- (-3,-3);
\draw[-]  (3,-3) -- (6,0) -- (3,3);
\draw[dashed] (-3,-3) -- (3,-3);
\draw[dashed] (-3,3) -- (3,3);
\draw[-] (-3,-3) -- (3,3);
\draw[-] (-3,3) -- (3,-3);
\filldraw[black] (3.25,1.5) circle (2pt);
\filldraw[black] (-3.25,1.5) circle (2pt);
\filldraw[black] (4,1.3) circle (2pt);
\filldraw[black] (-4,1.3) circle (2pt);
\node at (-4,1) {\footnotesize $2_L$};
\node at (-3.25,1.2) {\footnotesize $1_L$};
\node at (3.25,1.2) {\footnotesize $1_R$};
\node at (4,1) {\footnotesize $2_R$};
\draw[thick] (-3.25,1.5) to[out=-170,in=18] (-4,1.3);
\draw[thick] (3.25,1.5) to[out=-10,in=162] (4,1.3);
\node at (3.7,1.7) {\footnotesize $A_R$};
\node at (-3.7,1.7) {\footnotesize $A_L$};
\node at (3.3,3.3) {\footnotesize $w^+$};
\node at (-3.3,3.3) {\footnotesize $w^-$};
\draw[<->] (2.7,0.3) -- (3,0) -- (3.3,0.3);
\node at (2.4,0.6) {\footnotesize $x^-$};
\node at (3.6,0.6) {\footnotesize $x^+$};
\begin{scope}[yshift=0cm]
\draw[<->] (-2.7,-0.3) -- (-3,0) -- (-3.3,-0.3);
\node at (-2.4,-0.6) {\footnotesize $x^-$};
\node at (-3.6,-0.6) {\footnotesize $x^+$};
\end{scope}
%
%\draw[<->] (0.3,1.2) -- (0,1.5) -- (0.3,1.8);
%\node at (0.6,0.9) {\footnotesize $x^-$};
%\node at (0.6,2.1) {\footnotesize $x^+$};
%
\end{tikzpicture}
\caption{\footnotesize The Penrose diagram showing the eternal black hole. The bath regions $L$ and $R$ are half Minkowski spaces glued onto the central AdS space. There are two intervals $A_L$ and $A_R$ in each bath as shown.}
\label{fig1} 
\end{center}
\end{figure}
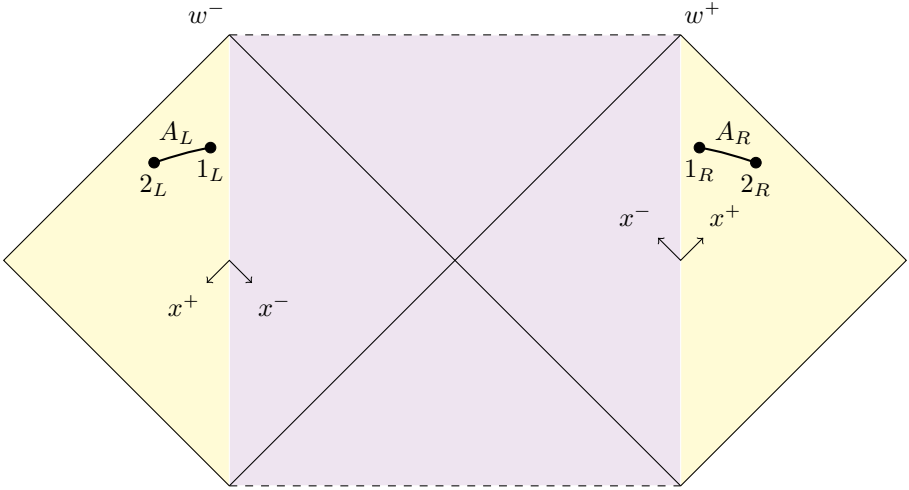

In the complementary gravity picture,  the Hartle-Hawking state and its (forward) time evolution is described by the two-sided black hole. The end-points of the intervals $A_L$ and $A_R$ will be labelled $2_L,1_L,1_R,2_R$ ordered along a Cauchy slice from left to right as shown in figure \ref{fig1}.
We define Kruskal-Szekeres (KS) coordinates $w^\pm$ that cover the whole of the spacetime, as well as null coordinates $x^\pm$ which are Minkowski coordinates in the baths and Schwarzschild coordinates in the AdS$_2$ region, related by
\EQ{
w^\pm=\pm e^{\pm2\pi x^\pm/\beta}\ .
\label{wxcoord}
}
The Schwarzschild time in the left region includes a sign change and an imaginary shift relative to the coordinate on the  right:
\EQ{
L:\quad x^\pm=-t\pm x+\frac{i\beta}2\ ,\qquad R:\quad x^\pm=t\pm x\ ,
\label{dsd}
}
where $x\geq0$ in both the left and right Minkowski baths.\footnote{It is natural that $x>0$ on both sides as $x$ is identified with the radial coordinate $r$ of the higher-dimensional black hole.} The imaginary shift is an important feature of the BCFT set up where we identify the two copies of the CFT's with the bath regions of the spacetime.

The AdS$_2$ black hole background (in JT gravity) in KS coordinates is described by the extended metric,
\be
ds^2=-\frac{4 dw^+ dw^-}{(1+w^+w^-)^2}\ ,
\ee
and the dilaton,
\be
\phi=\phi_0+\frac{2\pi\phi_r}{\beta}\frac{1-w^+w^-}{1+w^+w^-}\ .
\ee
The $L$ and $R$ horizons are at $w^+=0$ and $w^-=0$, whilst the AdS$_2$ boundaries are at $w^+w^-=-1$. The black hole has an entropy given by the dilaton value at the horizon since the dilaton is the area of the higher-dimensional, near extremal black hole,
\EQ{
S_\text{BH}^{(\beta)}=\frac{\phi(w^-=0)}{4G_N}=\frac{\pi c}{6\beta k}\ ,\qquad\text{where}\quad k\equiv\frac{G_Nc}{3\phi_r}\,.
}

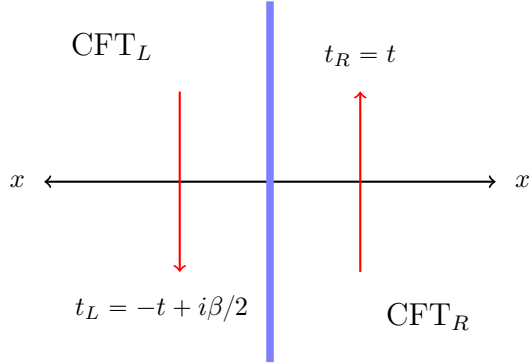
\begin{figure}[ht]
\begin{center}
\begin{tikzpicture}[scale=1.2]
\draw[thick,<->] (-2.5,0) -- (2.5,0);
\draw[blue!50,line width=1mm] (0,-2) -- (0,2);
\node at (2.8,0) {\footnotesize $x$};
\node at (-2.8,0) {\footnotesize $x$};
\node at (-1.75,1.5) {CFT$_L$};
\node at (1.75,-1.5) {CFT$_R$};
\draw[red,thick,->] (1,-1) -- (1,1);
\node at (1,1.4) {\footnotesize $t_R=t$};
\draw[red,thick,<-] (-1,-1) -- (-1,1);
\node at (-1.2,-1.4) {\footnotesize $t_L=-t+i\beta/2$};
\end{tikzpicture}
\caption{\footnotesize The BCFT set up with two half-Minkowski spaces $x\geq0$. There is also an imaginary shift which is naturally gives the TFD.}
\label{fig1.1} 
\end{center}
\end{figure}

In the BCFT picture, each copy of the CFT is defined on a half-Minkowski space that is isomorphic to the left or right bath region in the black hole geometry with Lorentzian null coordinates $x^\pm$ as in \eqref{dsd}: see figure \ref{fig1.1}. The imaginary shift on the left implements the necessary analytic continuation to obtain correlators in the TFD state from Euclidean thermal correlators \eqref{tfd}. 

\section{Entropy channels in BCFT}
\label{sec:bcft}

In this section, we calculate the entropies of the disjoint union of intervals in  a BCFT  prepared in the TFD state.
Let us consider two copies, $L$ and $R$, of the CFT in half-Minkowski spaces $x>0$ prepared in the TFD state, and whose time evolution can be followed by appropriate analytic continuation of Euclidean thermal correlators, 
\be
\big|\Psi_{\rm TFD}(t)\rangle=\sum_{n}e^{-2iE_nt}e^{-\beta E_n/2}|E_n\rangle_{L}\otimes|E_n\rangle_R\ .
\label{tfd}
\ee

\subsection{Semi-infinite intervals}

As a warm up exercise, we first look at two semi-infinite spatial intervals $A_L=[a_1,\infty]$ and $A_R=[a_2,\infty]$ in the free fermion theory. This situation was analysed from the perspective of the two-sided black hole with semi-infinite intervals in \cite{Almheiri:2019yqk, Almheiri:2019qdq}, from the holographic (large-$c$) BCFT viewpoint in \cite{Geng:2021iyq} and its explicit higher dimensional realizations \cite{Almheiri:2019psy}.\footnote{The authors of \cite{Geng:2021iyq} considered the entanglement entropy of the bipartition of a large-$c$ CFT on a strip i.e.~with {\em two} boundaries in the TFD state.} The example illustrates how the exact result at high temperatures is determined essentially by a competition between OPE channels.

R\'enyi entropies in CFT$_2$ are computed by correlation functions of twist and anti-twist operators, ${\cal T}_{n}$ and  $\overline{\cal T}_{n}$, which implement the replica trick in Euclidean signature \cite{Calabrese:2009qy}. For the present case, we need insertions of these operators at the (finite) endpoints of  $A_L$ and $A_R$. The correlators can then be analytically continued to Lorentzian signature, in which case we can write the real-time $n$-th R\'enyi entropy as the correlator
\be
S^{(n)}(A_L\cup A_R)\,=\,\frac{1}{1-n}\log\big\langle\overline{\cal T}_n (x^\pm_{1_L}) {\cal T}_n(x^\pm_{1_R})\big\rangle_{\rm BCFT}\ ,
\ee
with the coordinates defined in \eqref{dsd}. Evaluation of BCFT correlation functions is facilitated by the {\em doubling trick} which enables us to view them as correlators of a chiral CFT on the complex plane or the thermal cylinder \cite{Recknagel:2013uja, Sully:2020pza}:
\be
\big\langle\overline{\cal T}_{n}(x^\pm_{1_L}) {\cal T}_{n}(x^\pm_{1_R})\big\rangle_{\rm BCFT}=\big\langle{\cal T}_{n}(x^+_{1_L}){\overline{\cal T}}_{n}(x^-_{1_L}){\overline{\cal T}}_{n}(x^+_{1_R}){{\cal T}}_{n}(x^-_{1_R})\big\rangle\ .
\label{kgg}
\ee
The notation here requires some clarification: the twist operators on the left have conformal dimension $\Delta_n=c(n-1/n)/24$, whereas those on the right are in the chiral theory with half the conformal weight.
Since four-point functions are not fixed by conformal invariance alone, it is natural to focus on examples and/or physical regimes in which the results become tractable and potentially universal. We first focus on the free fermion theory for which exact results are known, and then argue that the high temperature features are generic, at least for theories that admit a free quasiparticle  description.
%This is analogous to the single interval example for the CFT on the two sided AdS-black hole coupled to CFT baths.

\paragraph{Free fermions:}for free fermions, the general expression for the R\'enyi entropy of a disjoint union of arbitrary number of intervals is known \cite{Casini:2009sr}. Viewing the free fermion BCFT as a chiral fermion theory on the double copy gives us the following expression for the 4-point function involving twist operator insertions \eqref{kgg}:
\EQ{
S^{(n)}(A_L\cup A_R)&=\frac{c(n+1)}{12n}\left(\log\sinh\tfrac\pi\beta(x^+_{1_L}-x^+_{1_R})\sinh\tfrac\pi\beta(x^-_{1_L}-x^-_{1_R}) +\log\eta\right)\ ,\\[8pt]
\text{where}&\qquad\eta=\frac{\sinh\tfrac\pi\beta(x^+_{1_L}-x^-_{1_L})\sinh\tfrac\pi\beta(x^+_{1_R}-x^-_{1_R})}{\sinh\tfrac\pi\beta(x^-_{1_R}-x^+_{1_L})\sinh\tfrac\pi\beta(x^-_{1_L}-x^+_{1_R})}\ .
}
Here, we have suppressed the additive logarithmic dependence on the UV cutoff.  
There is an additional finite contribution to the R\'enyi entropy in BCFT originating from the boundary degrees of freedom, namely the boundary entropy $s_b=\log g_b$ \cite{Friedan:2003yc}.
The free fermion system has $g_b=1$, so the boundary entropy is vanishing. The von Neumann entropy is then obtained by taking the limit $n\to1$. In the following, we will focus primarily on the von Neumann entropy.
\pgfdeclareimage[interpolate=true,width=8cm]{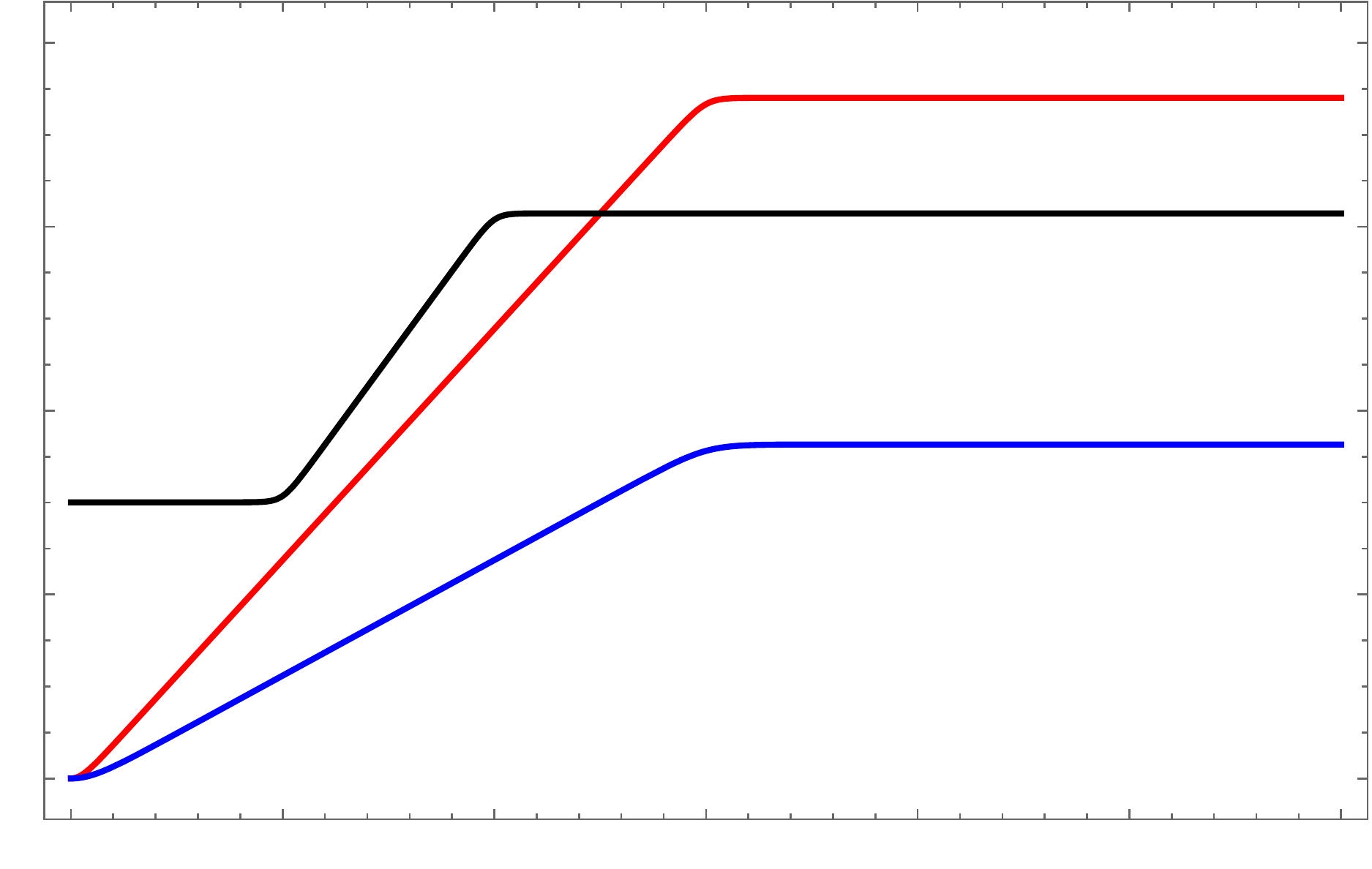}{pic3}
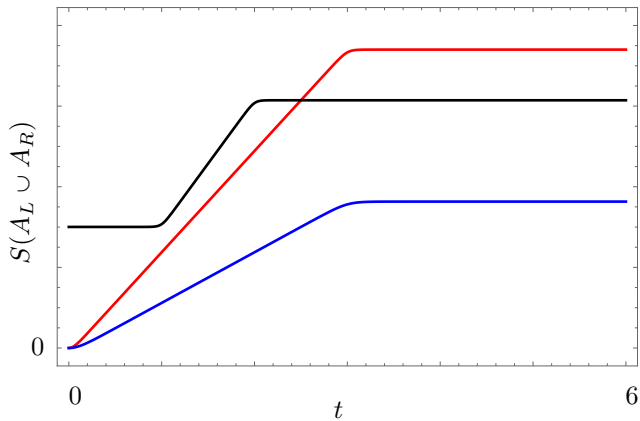
\begin{figure}[h]
\begin{center}
\begin{tikzpicture}[scale=1]
\pgftext[at=\pgfpoint{0cm}{0cm},left,base]{\pgfuseimage{pic3}} 
\node at (0.5,-0.1) {\footnotesize $0$};
\node at (7.9,-0.1) {\footnotesize $6$};
\node at (0,0.55) {\footnotesize$0$};
\node at (4,-0.3) {\footnotesize $t$};
\node[rotate=90] at (-0.2,2.6) {\footnotesize $S(A_L\cup A_R)$};
\end{tikzpicture}
\caption{\footnotesize The exact behaviour of  the von Neumann entropy for free fermion BCFT for semi-infinite intervals with $a_1=a_2=3$, $\beta=1$ (blue), $a_1=a_2=3$, $\beta=0.5$ (red) and $a_1=1$, $a_2=3$, $\beta=0.4$ (black).}
\label{fig2}
\end{center}
\end{figure}
Figure \ref{fig2} shows that in the symmetric case, the complete expression smoothly interpolates between early time linear growth and late time saturation, with the cross-ratio $\eta$ interpolating between $1$ and $0$. The transition between the regimes becomes sharper with increasing temperature. The early time linear growth is governed by the connected channel (cf.~figure \ref{fig3}) with $\eta=1$, reducing formally to the standard formula for the single interval entropy. 

The late time saturation ($\eta=0$), on the other hand, is due entirely to the disconnected channel which yields the product of the one-point functions of the twist and anti-twist fields in BCFT,\footnote{The one-point function for the twist field in the BCFT is given by the two-point correlator in the double copy chiral CFT,
\be
\big\langle{\cal T}_{n}(x^\pm)\big\rangle_{\rm BCFT}={g_b^{(1-n)}\epsilon^{\Delta_n}}/{\big(\sinh\tfrac{2\pi x}{\beta}\big)^{\Delta_n}}\,,\qquad \Delta_n=\tfrac{c}{24}\left(n-\tfrac1n\right)\,.
\ee
}
\be
\big\langle\overline{\cal T}_{n}(x^\pm_{1_L}) {\cal T}_{n}(x^\pm_{1_R})\big\rangle_{\rm BCFT}\Big|_{\eta\to 0}=\big\langle\overline{\cal T}_{n}(x^\pm_{1_L}) \big\rangle_{\rm BCFT}\,\big\langle{\cal T}_{n}(x^\pm_{1_R})\big\rangle_{\rm BCFT}\ .
\ee
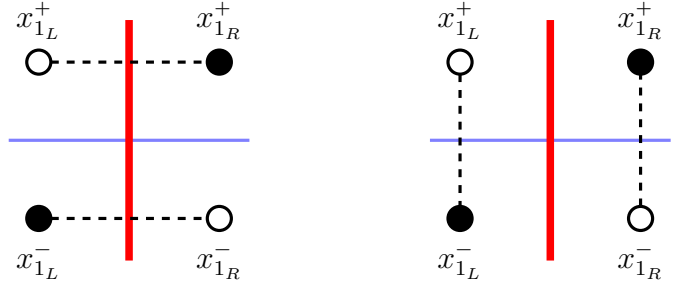
\begin{figure}[ht]
\begin{center}
\begin{tikzpicture}[scale=0.8]
\draw[blue!50,very thick] (-2,0) -- (2,0);
\draw[red,line width=1mm] (0,-2) -- (0,2);
\draw[very thick] (-1.5,1.3) circle (2mm);
\draw[very thick,fill=black] (1.5,1.3) circle (2mm);
\draw[very thick,fill=black] (-1.5,-1.3) circle (2mm);
\draw[very thick] (1.5,-1.3) circle (2mm);
\draw[very thick,dashed] (-1.3,1.3) -- (1.3,1.3);
\draw[very thick,dashed] (-1.3,-1.3) -- (1.3,-1.3);
\node at (-1.5,2) {$x^+_{1_L}$};
\node at (1.5,2) {$x^+_{1_R}$};
\node at (-1.5,-2) {$x^-_{1_L}$};
\node at (1.5,-2) {$x^-_{1_R}$};
\begin{scope}[xshift=7cm]
\draw[blue!50,very thick] (-2,0) -- (2,0);
\draw[red,line width=1mm] (0,-2) -- (0,2);
\draw[very thick] (-1.5,1.3) circle (2mm);
\draw[very thick,fill=black] (1.5,1.3) circle (2mm);
\draw[very thick,fill=black] (-1.5,-1.3) circle (2mm);
\draw[very thick] (1.5,-1.3) circle (2mm);
\draw[very thick,dashed] (-1.5,1.1) -- (-1.5,-1.1);
\draw[very thick,dashed] (1.5,-1.1) -- (1.5,1.1);
\node at (-1.5,2) {$x^+_{1_L}$};
\node at (1.5,2) {$x^+_{1_R}$};
\node at (-1.5,-2) {$x^-_{1_L}$};
\node at (1.5,-2) {$x^-_{1_R}$};
\end{scope}
\end{tikzpicture}
\caption{\footnotesize Schematic depiction of the two distinct channels of contraction for the  twist-anti-twist correlator in BCFT. The panel on the right yields the disconnected contribution given by the product of the one-point function of twist fields on the half-plane (in the Euclidean picture). Solid and hollow dots distinguish twist and anti-twist operator insertions.}
\label{fig3} 
\end{center}
\end{figure}
The behaviour of the cross-ratios and the correlator closely follows the expectation for  large-$c$ CFTs, where the two different regimes can be viewed as distinct saddle point approximations to  conformal blocks at large-$c$ \cite{Hartman:2013mia, Sully:2020pza}.\footnote{An argument for this follows from modular invariance which would suggest that  high temperature correlators should be determined by the vacuum conformal block, and the saddle point interpretation arises as a consequence of the thermodynamic limit instead of large-$c$.}
A useful lesson from this simple example is that the corrections to the two asymptotic behaviours at early and late times are exponentially small in the thermodynamic limit. For the symmetric situation $a_1=a_2\equiv a$, we get
\EQ{
\eta\,\approx\,\begin{cases}1-e^{-\frac{4\pi}{\beta}(a-t)}-e^{-\frac{4\pi}{\beta}(a+t)} - 2 e^{-\frac{4\pi}{\beta}a}\,,\qquad & t< a\ ,\\[5pt]
e^{-\frac{4\pi}{\beta}(t-a)}\left(1-e^{-\frac{4\pi}{\beta}(t-a)}-e^{-\frac{4\pi}{\beta}(a+t)} - 2 e^{-\frac{4\pi}{\beta}a}\right)\,,\qquad &t>a\ .
\end{cases}
}
The high temperature thermodynamic limit erases short range correlations and gives a universal classical result for interval entropies, but unitarity as required by Page's theorem makes its appearance via competition between two channels. 

In the high temperature regime, where free fermion and large-$c$ limits appear to  coincide in this example, the entropy formulas can  be written in terms of the thermodynamic entropy of the left/right moving radiation in an interval of null coordinate $[x^\pm_1,x^\pm_2]$; that is
\be
S_\text{rad}(x_1^\pm,x_2^\pm)\equiv\frac{\pi c}{6\beta} \,|x^\pm_1-x^\pm_2|\ .
\label{radlimit} 
\ee
The evolution of the high temperature entropy for the interval $A_L\cup A_R$ can written in terms of the thermodynamic entropy \eqref{radlimit} evaluated on the difference of null coordinates: 
\EQ{
S(A_L\cup A_R)&=\min\big(S_{\rm rad}(x^+_{1_L}, x^+_{1_R})+S_{\rm rad}(x^-_{1_L}, x^-_{1_R})\ ,\\ &\qquad\qquad S_{\rm rad}(x^+_{1_L}, x^-_{1_L})+S_{\rm rad}(x^+_{1_R}, x^-_{1_R}) + 2\log g_b\big)\ ,
}
where we have allowed for a non-zero boundary entropy $g_b$ and implemented the rule that every one-point function understood as a contraction between a point and its image is accompanied by the additive constant $\log g_b$. This yields the time evolution of the entropy in the high temperature limit,
\EQ{
S(A_L\cup A_R)&=\frac{\pi c}{6\beta} 
\min\big( \left|2t-(a_2-a_1)\right|+\left|2t+(a_2-a_1)\right|\ ,\\[5pt]
&\qquad\qquad\qquad 2(a_1+a_2) + \tfrac{12 \beta}{\pi c}\log g_b\big)\ .
\label{ucc}
}
 The expression reproduces the free fermion curve in figure \ref{fig2} with $g_b=1$. The two channels  exchange at the crossover time or the Page time,
 \be
 t_{\rm Page}=\frac{a_1+a_2}{2} +\frac{3\beta}{\pi c}\log g_b\,.
 \ee 
 
\subsection{Two finite intervals}

Now let us consider the two finite intervals, one in each CFT, not including the boundary as described in section \ref{s2}. To calculate the entropies in the TFD state we need the correlator of four twist and anti-twist operators on the semi-infinite thermal cylinder, appropriately continued to Lorentzian signature:
\be
C_{(4)}=\langle{\cal T}_n(x^\pm_{1_L}){\overline {\cal T}}_n(x^\pm_{2_L}){\overline{\cal T}}_n(x^\pm_{1_R}){\cal T}_n(x^\pm_{2_R})\rangle_{\rm BCFT}\,.
\ee
The doubling trick turns this into an 8-point chiral correlator on the infinite thermal cylinder,
\be
C_{(4)}=\langle{\cal T}_n(x^+_{1_L}){\overline{\cal T}}_n(x^-_{1_L}){\overline {\cal T}}_n(x^+_{2_L}){{\cal T}}_n(x^-_{2_L}){\overline{\cal T}}_n(x^+_{1_R}){{\cal T}}_n(x^-_{1_R}){\cal T}_n(x^+_{2_R}){\overline{\cal T}}_n(x^-_{2_R})\rangle\ .
\ee
There are several distinct OPE channels corresponding to pairs of twist-anti-twist operators potentially coming together.

\paragraph{Free fermions:}  the  BCFT correlator for the R\'enyi entropy can be obtained using the result of \cite{Casini:2009sr} (see also \cite{Reyes:2021npy}) after incorporating the doubling trick. The von Neumann entropy then follows in the $n\to1$ limit. The result is usefully written in terms of the mutual information 
  \be
  S(A_L\cup A_R)=S(A_L)+S(A_R)-I(A_L,A_R)\ .
  \ee
  Each interval has an entropy 
\EQ{
 S(A_{L})=\frac{c}{6}\log\frac{\ms(x^+_{2_L}-x^+_{1_L})\ms(x^-_{2_L}-x^-_{1_L})\ms(x^+_{1_L}-x^-_{1_L})\ms(x^+_{2_L}-x^-_{2_L})}{\ms(x^-_{2_L}-x^+_{1_L})\ms(x^-_{1_L}-x^+_{2_L})}\ ,
}
with a similar expression for $S(A_R)$. The mutual information, which encodes the cross-correlations between $A_L$ and $A_R$ is
\EQ{
I(A_L,A_R) =-\frac{c}{6}
\log &\Big\{\frac{\ms(x^+_{1_L}-x^+_{1_R})\ms(x^+_{2_L}-x^+_{2_R})\ms(x^-_{1_L}-x^-_{1_R})\ms(x^-_{2_L}-x^-_{2_R})}{\ms(x^+_{1_L}- x^+_{2_R})\ms(x^+_{2_L}- x^+_{1_R})\ms(x^-_{1_L}-x^-_{2_R})\ms(x^-_{2_L}-x^-_{1_R})}\\[8pt]
&\times\frac{\ms(x^+_{1_L}-x^-_{2_R})\ms(x^+_{2_L}-x^-_{1_R})\ms(x^+_{1_R}-x^-_{2_L})\ms(x^+_{2_R}-x^-_{1_L})}{\ms(x^+_{1_L}-x^-_{1_R})\ms(x^+_{2_L}- x^-_{2_R})\ms(x^+_{1_R}- x^-_{1_L})\ms(x^+_{2_R}- x^-_{2_L})}\Big\}\ .
}
Note that the imaginary shift \eqref{dsd} turns $\sinh\to\cosh$ in the mutual information. The entropy is plotted in figure \ref{fig5} for some indicative values.  
\pgfdeclareimage[interpolate=true,width=8cm]{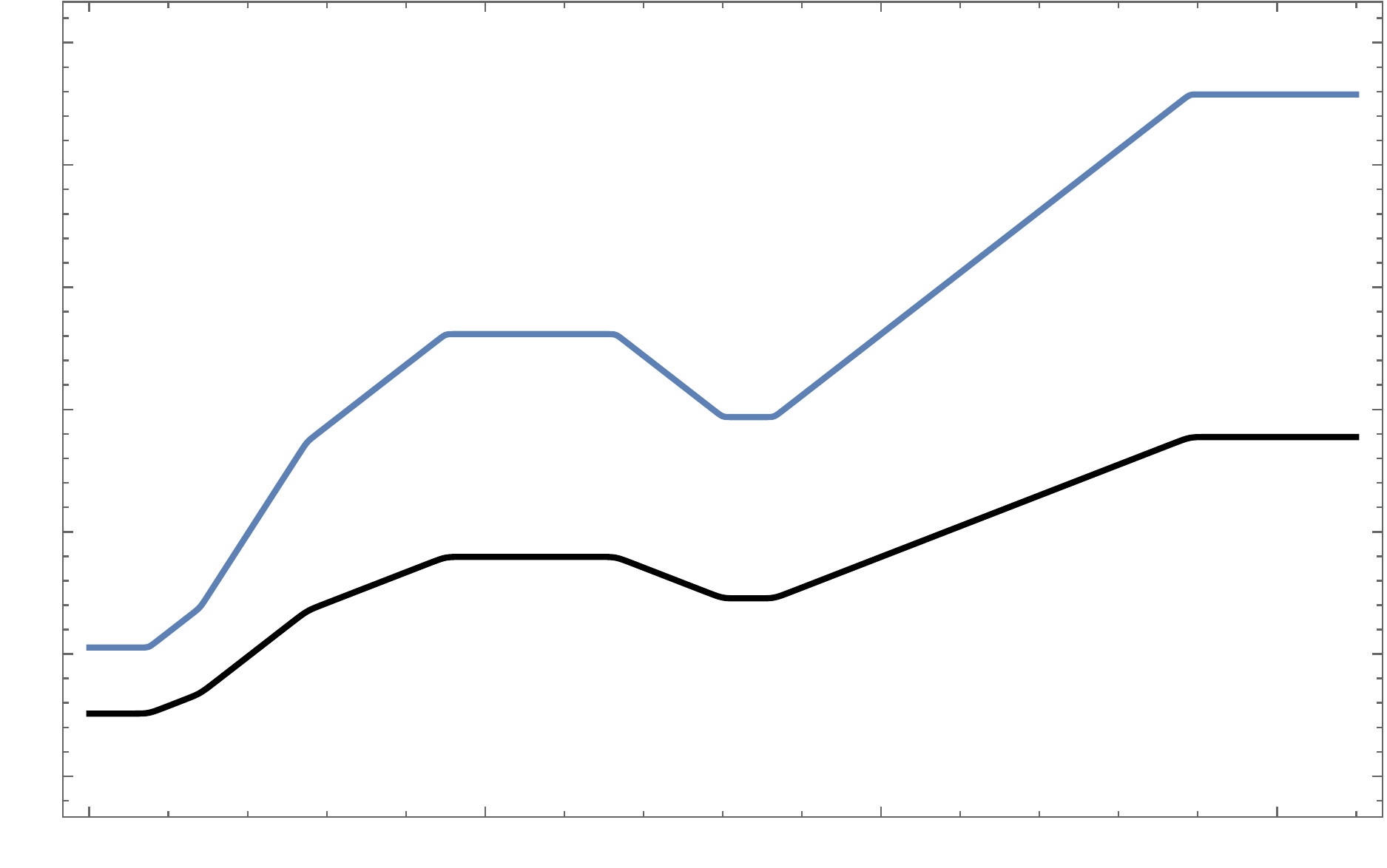}{pic2}
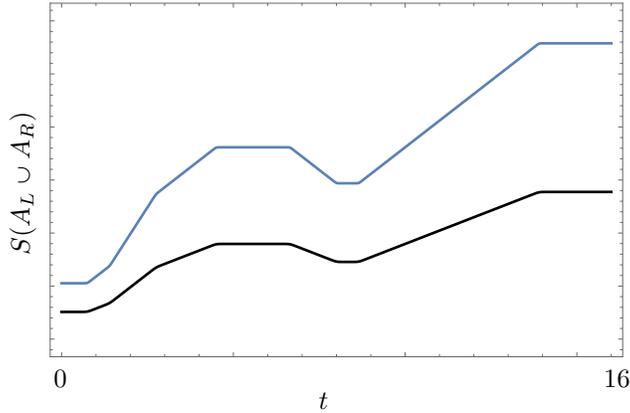
\begin{figure}[h]
\begin{center}
\begin{tikzpicture}[scale=1]
\pgftext[at=\pgfpoint{0cm}{0cm},left,base]{\pgfuseimage{pic2}} 
\node at (0.5,0) {\footnotesize $0$};
\node at (7.9,0) {\footnotesize $16$};
\node at (4,-0.3) {\footnotesize $t$};
\node[rotate=90] at (0,2.6) {\footnotesize $S(A_L\cup A_R)$};
\end{tikzpicture}
\caption{\footnotesize Exact result for free fermion entropy of the disjoint union of two intervals, one in each copy of the thermofield double state. The plot displayed is for $A_L=[2,12.5]$ and $A_R=[3.5,15.3]$ and for $\beta=0.25$ (blue) and $\beta=0.5$ (black). The dip in the entropy is clearly displayed.}
\label{fig5}
\end{center}
\end{figure}

\subsection{Large-$c$ OPE channels}

\begin{figure}[ht]
\begin{center}
\begin{tikzpicture}[scale=0.6]
\node at (-3.8,2.2) {\small(1)};
\draw[blue!50,very thick] (-3,0) -- (3,0);
\draw[red,line width=1mm] (0,-2) -- (0,2);
\draw[very thick] (-1,1.3) circle (1.5mm);
\draw[very thick,fill=black] (1,1.3) circle (1.5mm);
\draw[very thick,fill=black] (-1,-1.3) circle (1.5mm);
\draw[very thick] (1,-1.3) circle (1.5mm);
\draw[very thick] (-2.5,-1.3) circle (1.5mm);
\draw[very thick,fill=black] (2.5,-1.3) circle (1.5mm);
\draw[very thick,fill=black] (-2.5,1.3) circle (1.5mm);
\draw[very thick] (2.5,1.3) circle (1.5mm);
\draw[thick,dashed] (1.15,1.3) -- (2.35,1.3);
\draw[thick,dashed] (-1.15,-1.3) -- (-2.35,-1.3);
\draw[thick,dashed] (1.15,-1.3) -- (2.35,-1.3);
\draw[thick,dashed] (-1.15,1.3) -- (-2.35,1.3);
\node at (-1,2.2) {\footnotesize $x^+_{1_L}$};
\node at (1,2.2) {\footnotesize $x^+_{1_R}$};
\node at (-1,-2.2) {\footnotesize $x^-_{1_L}$};
\node at (1,-2.2) {\footnotesize $x^-_{1_R}$};
\node at (-2.5,2.2) {\footnotesize $x^+_{2_L}$};
\node at (2.5,2.2) {\footnotesize $x^+_{2_R}$};
\node at (-2.5,-2.2) {\footnotesize $x^-_{2_L}$};
\node at (2.5,-2.2) {\footnotesize $x^-_{2_R}$};
\begin{scope}[xshift=11cm]
\node at (-3.8,2.2) {\small(2)};
\draw[blue!50,very thick] (-3,0) -- (3,0);
\draw[red,line width=1mm] (0,-2) -- (0,2);
\draw[very thick] (-1,1.3) circle (1.5mm);
\draw[very thick,fill=black] (1,1.3) circle (1.5mm);
\draw[very thick,fill=black] (-1,-1.3) circle (1.5mm);
\draw[very thick] (1,-1.3) circle (1.5mm);
\draw[very thick] (-2.5,-1.3) circle (1.5mm);
\draw[very thick,fill=black] (2.5,-1.3) circle (1.5mm);
\draw[very thick,fill=black] (-2.5,1.3) circle (1.5mm);
\draw[very thick] (2.5,1.3) circle (1.5mm);
\draw[thick,dashed] (-0.85,1.3) -- (1.15,1.3);
\draw[thick,dashed] (-1.15,-1.3) -- (0.85,-1.3);
\draw[thick,dashed] (-2.35,-1.3) to[out=-15,in=-165] (2.35,-1.3);
\draw[thick,dashed] (-2.35,1.3) to[out=15,in=165] (2.35,1.3);
\node at (-1,2.2) {\footnotesize $x^+_{1_L}$};
\node at (1,2.2) {\footnotesize $x^+_{1_R}$};
\node at (-1,-2.2) {\footnotesize $x^-_{1_L}$};
\node at (1,-2.2) {\footnotesize $x^-_{1_R}$};
\node at (-2.5,2.2) {\footnotesize $x^+_{2_L}$};
\node at (2.5,2.2) {\footnotesize $x^+_{2_R}$};
\node at (-2.5,-2.2) {\footnotesize $x^-_{2_L}$};
\node at (2.5,-2.2) {\footnotesize $x^-_{2_R}$};
\end{scope}
\begin{scope}[yshift=-7cm]
\node at (-3.8,2.2) {\small(3)};
\draw[blue!50,very thick] (-3,0) -- (3,0);
\draw[red,line width=1mm] (0,-2) -- (0,2);
\draw[very thick] (-1,1.3) circle (1.5mm);
\draw[very thick,fill=black] (1,1.3) circle (1.5mm);
\draw[very thick,fill=black] (-1,-1.3) circle (1.5mm);
\draw[very thick] (1,-1.3) circle (1.5mm);
\draw[very thick] (-2.5,-1.3) circle (1.5mm);
\draw[very thick,fill=black] (2.5,-1.3) circle (1.5mm);
\draw[very thick,fill=black] (-2.5,1.3) circle (1.5mm);
\draw[very thick] (2.5,1.3) circle (1.5mm);
\draw[thick,dashed] (1,1.15) -- (1,-1.15);
\draw[thick,dashed] (-1,1.15) -- (-1,-1.15);
\draw[thick,dashed] (-2.35,-1.3) to[out=-15,in=-165] (2.35,-1.3);
\draw[thick,dashed] (-2.35,1.3) to[out=15,in=165] (2.35,1.3);
\node at (-1,2.2) {\footnotesize $x^+_{1_L}$};
\node at (1,2.2) {\footnotesize $x^+_{1_R}$};
\node at (-1,-2.2) {\footnotesize $x^-_{1_L}$};
\node at (1,-2.2) {\footnotesize $x^-_{1_R}$};
\node at (-2.5,2.2) {\footnotesize $x^+_{2_L}$};
\node at (2.5,2.2) {\footnotesize $x^+_{2_R}$};
\node at (-2.5,-2.2) {\footnotesize $x^-_{2_L}$};
\node at (2.5,-2.2) {\footnotesize $x^-_{2_R}$};
\end{scope}
\begin{scope}[yshift=-7cm,xshift=11cm]
\node at (-3.8,2.2) {\small(4)};
\draw[blue!50,very thick] (-3,0) -- (3,0);
\draw[red,line width=1mm] (0,-2) -- (0,2);
\draw[very thick] (-1,1.3) circle (1.5mm);
\draw[very thick,fill=black] (1,1.3) circle (1.5mm);
\draw[very thick,fill=black] (-1,-1.3) circle (1.5mm);
\draw[very thick] (1,-1.3) circle (1.5mm);
\draw[very thick] (-2.5,-1.3) circle (1.5mm);
\draw[very thick,fill=black] (2.5,-1.3) circle (1.5mm);
\draw[very thick,fill=black] (-2.5,1.3) circle (1.5mm);
\draw[very thick] (2.5,1.3) circle (1.5mm);
\draw[thick,dashed] (0.9,1.3) -- (2.4,1.3);
\draw[thick,dashed] (2.4,-1.2) -- (-0.9,1.3);
\draw[thick,dashed] (-1.15,-1.3) -- (-2.35,-1.3);
\draw[thick,dashed] (0.9,-1.2) -- (-2.5,1.3);
\node at (-1,2.2) {\footnotesize $x^+_{1_L}$};
\node at (1,2.2) {\footnotesize $x^+_{1_R}$};
\node at (-1,-2.2) {\footnotesize $x^-_{1_L}$};
\node at (1,-2.2) {\footnotesize $x^-_{1_R}$};
\node at (-2.5,2.2) {\footnotesize $x^+_{2_L}$};
\node at (2.5,2.2) {\footnotesize $x^+_{2_R}$};
\node at (-2.5,-2.2) {\footnotesize $x^-_{2_L}$};
\node at (2.5,-2.2) {\footnotesize $x^-_{2_R}$};
\end{scope}
\end{tikzpicture}
\caption{\footnotesize Examples of competing channels for the two-interval R\'enyi entropy in the TFD state. The last two figures illustrate partially disconnected channels that are associated with island contributions in the black hole case. }
\label{fig4} 
\end{center}
\end{figure}
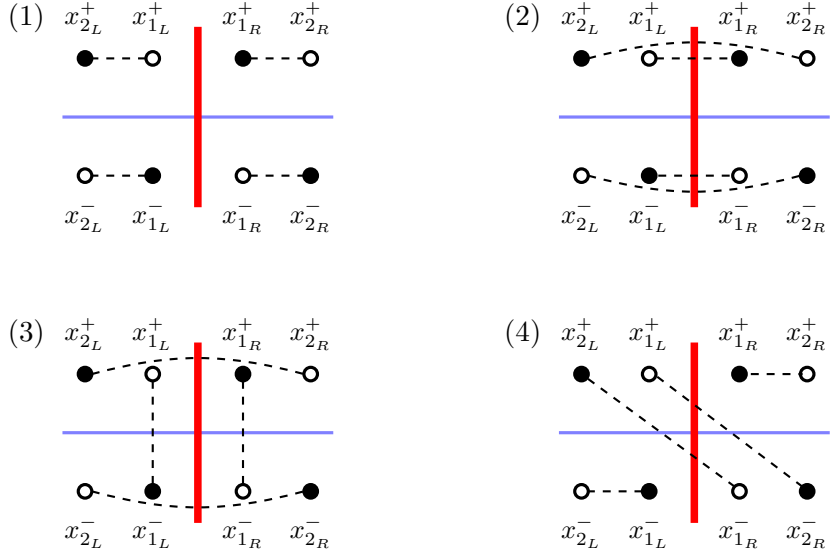

The high temperature evolution of $S(A_L\cup A_R)$ for free fermions follows a curve that becomes more and more of a piecewise linear function of time as $\beta$ decreases: see figure \ref{fig5}. This behaviour is generic at high temperatures and can be understood as a  competition between different OPE channels of the twist field correlators. Some of the channels/singularities that we encounter here are not expected to appear in non-rational CFTs or in  holographic large-$c$ BCFTs \cite{Asplund:2015eha}.
There are multiple channels for factorization of the R\'enyi entropy correlators. These can be classified according to  the distinct ways in which four twist operators can pair up with four anti-twist operators. This gives us $4!=24$ possible channels. Of these, most tend to have high entropy and so be unfavourable  and only a small number of the channels can actually compete as shown in figures \ref{fig4} and \ref{fig8}.

\paragraph{Two symmetric intervals:} for example, when the two intervals $A_L$ and $A_R$ are taken to be identical copies of each other, $a_1=a_2= a$ and $b_1=b_2=b$, the four relevant channels are illustrated in figure \ref{fig4}. 

The high temperature entropy is then given by
\EQ{
S(A_L\cup A_R)=\frac{2\pi c }{3\beta}&\min\big(b-a\,,\,2t\,, \,t+ a+\tfrac{3\beta}{\pi c}\log g_b\ ,\\[5pt]
&\tfrac12(b-a)+ \left|t-\tfrac12(a+b)\right|+\tfrac{3\beta}{\pi c}\log g_b\big)\ .
\label{BCFTentropy}
}
In general, there is a dependence on the value of the boundary entropy $\log g_b$, and this should be set to zero for the free fermion theory.

When the interval length is large, or more precisely $b >3a$, we can identify five distinct regimes (I)-(V) in the evolution of the entropy, as displayed in figure \ref{fig6}. The early time linear growth region (I) is controlled by the connected channel $(2)$ shown in figure \ref{fig4}, until there is a transition to the channel $(3)$ with a lower slope at
\be
t_{2\to 3}=a\,+\,\frac{3\beta}{\pi c}\,\log g_b\,.
\ee
In regime (II), dominated by channel (3), there is a partial factorization of the twist operator correlation function. A second transition between  two distinct channels $(3)$ and $(4)$ occurs at
\be
t_{3\to 4}=\frac{b-a}2\ .
\ee
This brings us to region (III) described by the channel (4), which is particularly interesting as the entropies of entanglement now experience a dip at
\be 
t_{\rm dip}=\frac{a+b}2\,.
\ee
This is a non-trivial feature that we will explain in detail in sections \ref{s5} and \ref{s6}.  At this juncture we  note  that a decrease in the entropy indicates a partial purification of certain modes in $A_R$ by partner modes in $A_L$. 
At the minimum, in the free fermion theory $(g_b=1)$, the entropy is equal to one half of the saturation value at large $t$ which is just the entropy of a single interval:
\be
S(A_L\cup A_R)\big|_{t=t_\text{dip}}= S(A_L)=S(A_R)\,.
\ee 
Later, in section \ref{s6}, we will discuss why this happens.
For $t>t_{\rm dip}$, the entropy in channel $(4)$ continues to grow until it saturates at the thermal value after a final, third exchange of channels  at 
\be
t_{4\to 1}=b-\frac{3\beta}{\pi c}\,\log g_b\ ,
\ee
following which the entropies of the intervals $A_L$ and $A_R$ saturate at their thermal values. The competition between channels precisely reproduces the high temperature result in the free fermion theory with $g_b=1$.
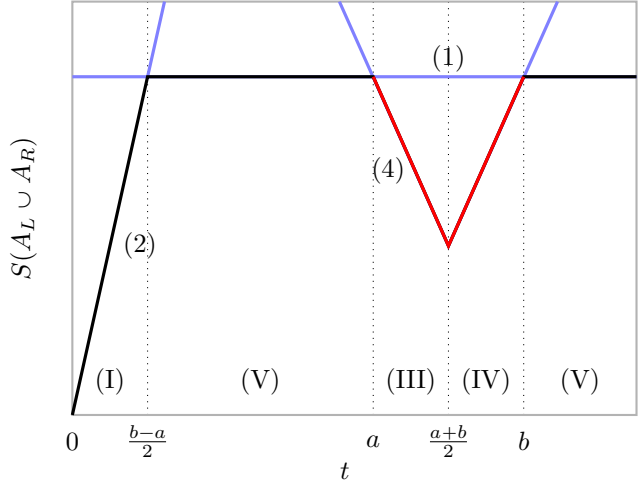
\begin{figure}[h]
\begin{center}
\begin{tikzpicture} [scale=0.5]
\draw[dotted] (2,0) -- (2,11);
\draw[dotted] (8,0) -- (8,11);
\draw[dotted] (10,0) -- (10,11);
\draw[dotted] (12,0) -- (12,11);
\draw[thick,black!30] (0,0) -- (0,11) -- (15,11) -- (15,0) --cycle;
%
%\draw[very thick,black!30] (0,4.5) -- (2.9,11);
%
\draw[very thick,blue!50] (2,9) -- (2.45,11);
\draw[very thick,blue!50] (7.1,11) -- (10,4.5) -- (12.9,11);
\draw[very thick,blue!50] (0,9) -- (15,9);
\draw[very thick] (0,0) -- (2,9) -- (8,9) -- (10,4.5) -- (12,9) -- (15,9);
\draw[very thick,red] (8,9) -- (10,4.5) -- (12,9);
\node at (1,0.9) {\footnotesize(I)};
\node at (5,0.9) {\footnotesize(V)};
\node at (9,0.9) {\footnotesize(III)};
\node at (11,0.9) {\footnotesize(IV)};
\node at (13.5,0.9) {\footnotesize(V)};
\node at (0,-0.7) {\footnotesize0};
\node at (8,-0.7) {\footnotesize$a$};
\node at (2,-0.7) {\footnotesize$\tfrac{b-a}2$};
\node at (10,-0.7) {\footnotesize$\tfrac{a+b}2$};
\node at (12,-0.7) {\footnotesize$b$};
\node at (7.25,-1.5) {\footnotesize $t$};
\node[rotate=90] at (-1.3,5.5) {\footnotesize $S(A_L\cup A_R)$};
\node at (10,9.5) {\footnotesize (1)};
\node at (1.8,4.5) {\footnotesize (2)};
\node at (8.4,6.5) {\footnotesize (4)};
\end{tikzpicture}
\caption{\footnotesize Evolution of entropy for two small  intervals $3a>b$ with $a=7$ and $b=8$ shown. 
The numbers (1), etc., refer to the channels that contribute. The system appears to equilibrate, however, there is a later dip in the entropy corresponding to the island saddle in the BH case.}
\label{fig7}
\end{center}
\end{figure}

For small interval lengths, precisely $b<3a$, and small $\log g_b$ we get a different scenario as shown in figure \ref{fig7} in which the appearance of the dip is arguably more striking. In this scenario, channel (3)  remains subdominant at all times. Instead we have a direct transition from the initial growth phase to thermal equilibrium at
\be
t_{2\to 1}= \frac{b-a}2\ .
\ee
But the equilibrium is disturbed by an exchange of channels from (1) to (4) at 
\be
t_{1\to 4}=a+\frac{3\beta} {\pi c}\log g_b\,.
\ee
The reason for this is the presence of the CFT boundary: modes reflected off the $R$ boundary and their purifiers in CFT$_L$ enter $A_L\cup A_R$. 
The ensuing dip is symmetric, turning around at $t_\text{dip}=\frac12(a+b)$ and again reaching equilibrium at $t_{4\to 1}$. 

\begin{figure}[h]
\begin{center}
\begin{tikzpicture}[scale=0.6]
\node at (-3.8,2.2) {\small(5)};
\draw[blue!50,very thick] (-3,0) -- (3,0);
\draw[red,line width=1mm] (0,-2) -- (0,2);
\draw[very thick] (-1,1.3) circle (1.5mm);
\draw[very thick,fill=black] (1,1.3) circle (1.5mm);
\draw[very thick,fill=black] (-1,-1.3) circle (1.5mm);
\draw[very thick] (1,-1.3) circle (1.5mm);
\draw[very thick] (-2.5,-1.3) circle (1.5mm);
\draw[very thick,fill=black] (2.5,-1.3) circle (1.5mm);
\draw[very thick,fill=black] (-2.5,1.3) circle (1.5mm);
\draw[very thick] (2.5,1.3) circle (1.5mm);
\draw[thick,dashed] (-2.35,1.3) to[out=15,in=165] (2.35,1.3);
\draw[thick,dashed] (1,1.15) -- (1,-1.15);
\draw[thick,dashed] (-1.15,-1.3) -- (-2.35,-1.3);
\draw[thick,dashed] (-0.9,1.2) -- (2.4,-1.2);
\node at (-1,2.2) {\footnotesize $x^+_{1_L}$};
\node at (1,2.2) {\footnotesize $x^+_{1_R}$};
\node at (-1,-2.2) {\footnotesize $x^-_{1_L}$};
\node at (1,-2.2) {\footnotesize $x^-_{1_R}$};
\node at (-2.5,2.2) {\footnotesize $x^+_{2_L}$};
\node at (2.5,2.2) {\footnotesize $x^+_{2_R}$};
\node at (-2.5,-2.2) {\footnotesize $x^-_{2_L}$};
\node at (2.5,-2.2) {\footnotesize $x^-_{2_R}$};
\begin{scope}[xshift=11cm]
\node at (-3.8,2.2) {\small(6)};
\draw[blue!50,very thick] (-3,0) -- (3,0);
\draw[red,line width=1mm] (0,-2) -- (0,2);
\draw[very thick] (-1,1.3) circle (1.5mm);
\draw[very thick,fill=black] (1,1.3) circle (1.5mm);
\draw[very thick,fill=black] (-1,-1.3) circle (1.5mm);
\draw[very thick] (1,-1.3) circle (1.5mm);
\draw[very thick] (-2.5,-1.3) circle (1.5mm);
\draw[very thick,fill=black] (2.5,-1.3) circle (1.5mm);
\draw[very thick,fill=black] (-2.5,1.3) circle (1.5mm);
\draw[very thick] (2.5,1.3) circle (1.5mm);
\draw[thick,dashed] (1.15,1.3) -- (2.35,1.3);
\draw[thick,dashed] (-1.15,1.3) -- (-2.35,1.3);
\draw[thick,dashed] (-1.15,-1.3) -- (0.85,-1.3);
\draw[thick,dashed] (-2.35,-1.3) to[out=-15,in=-165] (2.35,-1.3);
\node at (-1,2.2) {\footnotesize $x^+_{1_L}$};
\node at (1,2.2) {\footnotesize $x^+_{1_R}$};
\node at (-1,-2.2) {\footnotesize $x^-_{1_L}$};
\node at (1,-2.2) {\footnotesize $x^-_{1_R}$};
\node at (-2.5,2.2) {\footnotesize $x^+_{2_L}$};
\node at (2.5,2.2) {\footnotesize $x^+_{2_R}$};
\node at (-2.5,-2.2) {\footnotesize $x^-_{2_L}$};
\node at (2.5,-2.2) {\footnotesize $x^-_{2_R}$};
\end{scope}
\begin{scope}[yshift=-7cm]
\node at (-3.8,2.2) {\small(7)};
\draw[blue!50,very thick] (-3,0) -- (3,0);
\draw[red,line width=1mm] (0,-2) -- (0,2);
\draw[very thick] (-1,1.3) circle (1.5mm);
\draw[very thick,fill=black] (1,1.3) circle (1.5mm);
\draw[very thick,fill=black] (-1,-1.3) circle (1.5mm);
\draw[very thick] (1,-1.3) circle (1.5mm);
\draw[very thick] (-2.5,-1.3) circle (1.5mm);
\draw[very thick,fill=black] (2.5,-1.3) circle (1.5mm);
\draw[very thick,fill=black] (-2.5,1.3) circle (1.5mm);
\draw[very thick] (2.5,1.3) circle (1.5mm);
\draw[thick,dashed] (-1,1.15) -- (-1,-1.15);
\draw[thick,dashed] (1.15,1.3) -- (2.35,1.3);
\draw[thick,dashed] (-2.35,-1.3) to[out=-15,in=-165] (2.35,-1.3);
\draw[thick,dashed] (-2.4,1.2) -- (0.9,-1.2);
\node at (-1,2.2) {\footnotesize $x^+_{1_L}$};
\node at (1,2.2) {\footnotesize $x^+_{1_R}$};
\node at (-1,-2.2) {\footnotesize $x^-_{1_L}$};
\node at (1,-2.2) {\footnotesize $x^-_{1_R}$};
\node at (-2.5,2.2) {\footnotesize $x^+_{2_L}$};
\node at (2.5,2.2) {\footnotesize $x^+_{2_R}$};
\node at (-2.5,-2.2) {\footnotesize $x^-_{2_L}$};
\node at (2.5,-2.2) {\footnotesize $x^-_{2_R}$};
\end{scope}
\begin{scope}[yshift=-7cm,xshift=11cm]
\node at (-3.8,2.2) {\small(8)};
\draw[blue!50,very thick] (-3,0) -- (3,0);
\draw[red,line width=1mm] (0,-2) -- (0,2);
\draw[very thick] (-1,1.3) circle (1.5mm);
\draw[very thick,fill=black] (1,1.3) circle (1.5mm);
\draw[very thick,fill=black] (-1,-1.3) circle (1.5mm);
\draw[very thick] (1,-1.3) circle (1.5mm);
\draw[very thick] (-2.5,-1.3) circle (1.5mm);
\draw[very thick,fill=black] (2.5,-1.3) circle (1.5mm);
\draw[very thick,fill=black] (-2.5,1.3) circle (1.5mm);
\draw[very thick] (2.5,1.3) circle (1.5mm);
\draw[thick,dashed] (-0.85,1.3) -- (1.15,1.3);
\draw[thick,dashed] (-2.35,1.3) to[out=15,in=165] (2.35,1.3);
\draw[thick,dashed] (1.15,-1.3) -- (2.35,-1.3);
\draw[thick,dashed] (-1.15,-1.3) -- (-2.35,-1.3);
\node at (-1,2.2) {\footnotesize $x^+_{1_L}$};
\node at (1,2.2) {\footnotesize $x^+_{1_R}$};
\node at (-1,-2.2) {\footnotesize $x^-_{1_L}$};
\node at (1,-2.2) {\footnotesize $x^-_{1_R}$};
\node at (-2.5,2.2) {\footnotesize $x^+_{2_L}$};
\node at (2.5,2.2) {\footnotesize $x^+_{2_R}$};
\node at (-2.5,-2.2) {\footnotesize $x^-_{2_L}$};
\node at (2.5,-2.2) {\footnotesize $x^-_{2_R}$};
\end{scope}
\end{tikzpicture}
\caption{\footnotesize Some additional channels contributing to the generic two-interval R\'enyi entropies in the TFD state.}
\label{fig8}
\end{center}
\end{figure}
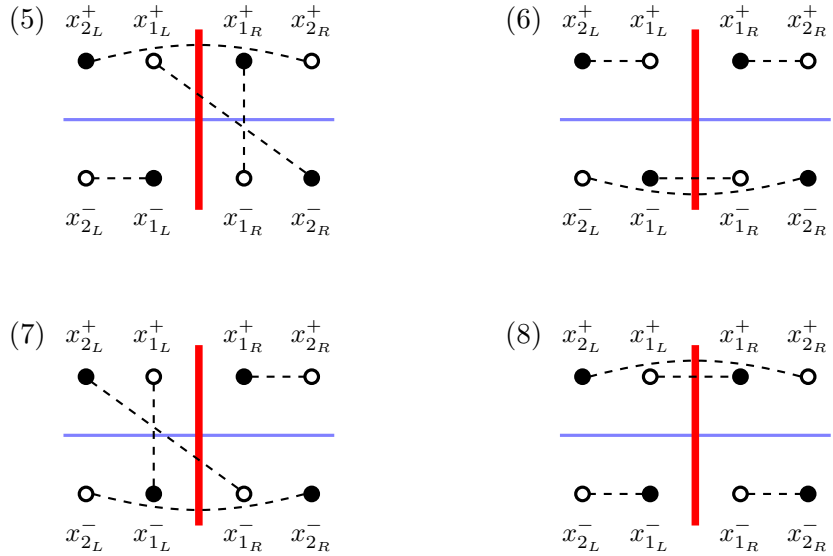

\begin{figure}[h]
\begin{center}
\begin{tikzpicture} [scale=0.6]
\draw[dotted] (2.75,0) -- (2.75,10);
\draw[dotted] (4.5,0) -- (4.5,10);
\draw[dotted] (6.65,0) -- (6.65,10);
\draw[dotted] (8,0) -- (8,10);
\draw[dotted] (8.65,0) -- (8.65,10);
\draw[dotted] (13.9,0) -- (13.9,10);
\draw[dotted] (2.75,0) -- (2.75,10);
\node at (1.38,0.9) {\footnotesize(I)};
\node at (3.63,0.9) {\footnotesize(II)};
\node at (5.58,0.9) {\footnotesize(IIA)};
\node at (7.33,0.9) {\footnotesize(III)};
\node at (8.33,1.7) {\footnotesize(IIIA)};
\node at (11.23,0.9) {\footnotesize(IV)};
\node at (14.45,0.9) {\footnotesize(V)};
%
%\draw[very thick,black!30] (0,6.04) -- (5.68,10);
%
\draw[very thick,blue!50] (1.4,1.95) -- (7.17,10);
\draw[very thick,blue!50] (0,9.69) -- (8,4.12);
\draw[very thick,blue!50] (15,8.55) -- (8.65,4.12);
\draw[very thick,blue!50] (0,6.04) -- (1.4,5.06) -- (8,5.06) -- (15,9.94); 
\draw[very thick,blue!50] (0,7.78) -- (15,7.78); 
\draw[very thick,blue!50] (0,2.9) -- (1.4,2.9) -- (2.75,3.84) -- (11.59,10); 
\draw[thick,black!30] (0,0) -- (0,10) -- (15,10) -- (15,0) --cycle;
\draw[very thick] (0,1.50) -- (0.75,1.50) -- (1.4,1.95) -- (2.75,3.84);
\draw[red,very thick] (2.75,3.84) -- (4.5,5.06) -- (6.65,5.06) -- (8,4.12) -- (8.65,4.12) -- (13.9,7.78);
\draw[very thick] (13.9,7.78) -- (15,7.78);
\node at (7.5,-1.5) {\footnotesize $t$};
\node at (0,-0.5) {\footnotesize $0$};
%\node at (15,-0.5) {\footnotesize $15$};
\node[rotate=90] at (-1.3,5) {\footnotesize $S(A_L\cup A_R)$};
\node at (10,8.2) {\footnotesize (1)};
\node at (9.2,6.5) {\footnotesize (5)};
\node at (2.3,2.3) {\footnotesize (2)};
\node at (1,3.4) {\footnotesize (3)};
\node at (9.5,4) {\footnotesize (4)};
\node at (2.7,-0.7) {\footnotesize$\tfrac{a_1+a_2}2$};
\node at (4.5,-0.7) {\footnotesize$\tfrac{b_1-a_2}2$};
\node at (6.3,-0.7) {\footnotesize$\tfrac{b_2-a_1}2$};
\node at (7.9,-0.7) {\footnotesize$\tfrac{a_2+b_1}2$};
\node at (9.3,-0.7) {\footnotesize$\tfrac{a_1+b_2}2$};
\node at (14,-0.7) {\footnotesize$\tfrac{b_1+b_2}2$};
\end{tikzpicture}
\caption{\footnotesize Plot of the entropy in the high temperature limit with non-symmetric intervals $A_L = [2,12.5]$ and $A_R = [3.5,15.3]$. As in previous plots, channels are shown in blue and labelled with a number $(1)$, etc. The red portion corresponds to the island saddle for the black hole.}
\label{fig9}
\end{center}
\end{figure}
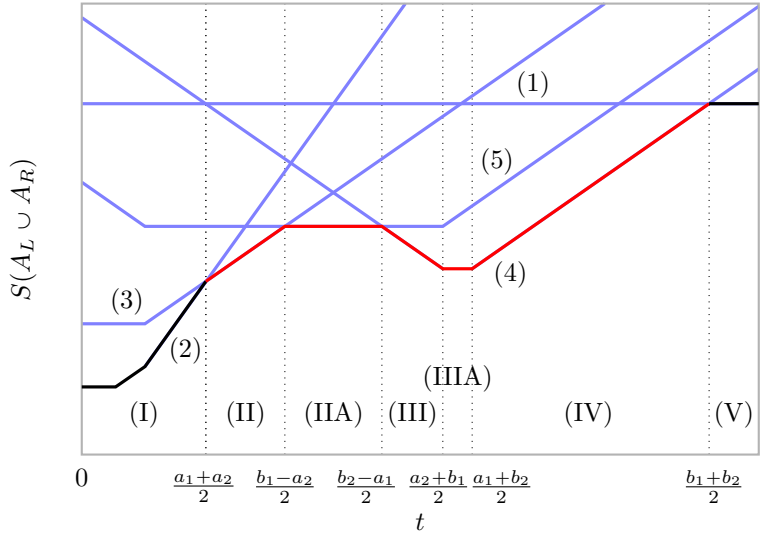

\paragraph{Generic two-interval case:} when the intervals $A_L$ and $A_R$ are chosen generically, we find that additional channels (figure \ref{fig8}) come into play alongside those depicted in figure \ref{fig4}. The effect of these additional contributions is shown in figure \ref{fig9}, where the crossover between channels (3) and (4) is flattened  by the appearance of channel (5), and the minimum point of the dip in channel (4) also appears flattened out.
%For the specific intervals indicated in figure \ref{fig7} (right panel),  channel $(5)$  above joins the set of competing saddles.
 By varying the  interval lengths and positions of the endpoints one can see various subsets of connected (no-island) and disconnected (island) channels becoming dominant during different intermediate time regimes, while early time growth and late time saturation are always  determined by the connected (no-island) OPE channels (2) and (1) respectively.
 
In the next section, we will examine how the different BCFT channels (of the free fermion theory) manifest themselves in the effective gravitational description of the free CFT coupled in the black hole background.

\section{Entropy saddles in JT gravity}
\label{sec:gravity}

The entropy of subsets $A$ of Hawking radiation in the bath region, as measured by asymptotic observers at $\mathscr{I}^+$, receive contributions from semiclassical replica wormhole saddle points of the gravitational path integral \cite{Penington:2019kki, Almheiri:2019qdq}. These yield QES, points in $1+1$ dimensions, the boundaries of additional intervals $I$, the islands,\footnote{Strictly speaking the island is the domain of dependence of any Cauchy surface that joins a pair of QES.} with positions determined by extremizing the generalized entropy functional for the von Neumann entropy of $A$,
\be
S_{\rm gen}(A)=\underset{I}\ext\Big\{\sum_{\partial I}\frac{{\rm Area}(\partial I)}{4G_N}+S_{\rm CFT}(I\cup A)\Big\}\ .
\label{pii}
\ee
In JT gravity the area term is the value of the dilaton evaluated at the QES.

We expect that the holographic equivalent of the generalized entropy is the BCFT entanglement entropy.
Non-trivial islands correspond to disconnected contributions to BCFT correlators. We expect the number of QES  to be determined by the number of twist operators contracted with their BCFT images in the relevant BCFT correlator. The area term in $S_{\rm gen}$ corresponds to the boundary entropy in the BCFT language. It encodes contributions that are not captured by the no-island or Hawking, saddle $S_\varnothing(A)$, which is the naive entropy of the CFT degrees of freedom in the bath. In the BCFT picture, this is given by the fully connected OPE channels for twist field correlators.

\subsection{Semi-infinite intervals}
The case with semi-infinite intervals was considered in \cite{Almheiri:2019qdq}. The entropy grows linearly initially corresponding to the no-island, or Hawking saddle. At the Page time, the entropy saturates as an island saddle takes over.  
At very late times, there are a pair of QES outside the horizon. Interestingly, careful examination reveals that the QES begin life {\em inside} the horizon and migrate out at an intermediate time scale. For two symmetric semi-infinite intervals, stretching from $x=a$ to $\infty$ in the $L$ and $R$ baths, the KS coordinates of the endpoints in the bath  are $w_{1_L}^\mp=w_{1_R}^\pm= \pm e^{2\pi(\pm t+a)/\beta}$. Then the locations of the QES (always staying close to the horizon) are given by, 
\EQ{
w_{1_Q}^-=-\frac s{w_{1_L}^+}\ ,\qquad w_{1_Q}^+=-\frac s{w_{1_L}^-} - \frac{1}{s}w_{1_R}^-\left(w_{1_L}^+\right)^2\ , \qquad w_{2_Q}^\pm=w_{1_Q}^\mp
\label{QESinf}
}
where we have defined $s=\beta k/2\pi\ll1$. Hence the QES exit their respective horizons at 
\be
t_{\rm exit}\,=\,2a+\frac\beta{2\pi}\ln \frac{2\pi}{\beta k}\,,
\ee
where the second term in the expression is the scrambling time, which we will take to be small in the discussion below.
For large enough intervals this is bigger than the Page time,
\be
t_{\rm Page}= a+ \frac{3\beta}{\pi c}S_{\rm BH}^{(\beta)}
\ee
after which the entropy saturates.
At high temperature the entropy is precisely as in \eqref{ucc} but with $\log g_b$ replaced by the Bekenstein-Hawking entropy $S_\text{BH}^{(\beta)}$. 

\subsection{No-island (Hawking) saddle }

We first discuss the Hawking saddle for the von Neumann entropy of $A_L\cup A_R$ where for simplicity the intervals are chosen symmetrically in the left and right baths $a_1=a_2\equiv a$ and $b_1=b_2\equiv b$. As in section \ref{sec:bcft}, we begin by examining large intervals i.e.~those with $b>3a$. For these, motivated by the BCFT analysis summarized in figure \ref{fig6}, we separately analyse each temporal regime,
\EQ{
\text{(I)}&=\{0< t< a\}\ ,\qquad\qquad\qquad\qquad \text{(II)}=\{a< t< \tfrac12(b-a)\}\ ,\\[5pt] \text{(III)}&=\{\tfrac12(b-a)< t<\tfrac12(a+b)\}\ ,\qquad \text{(IV)}=\{\tfrac12(a+b)< t< b\}\ ,\\[5pt] \text{(V)}&=\{b< t\}\ .
\label{nbc}
}
We will need to exploit the high temperature limit in order find the QES for the island saddle.

For the no-island saddle, there are four points in the bath, namely the endpoints of $A_L$ and $A_R$ whose lightcone coordinates $x^\pm$ determine the ordering of the magnitudes of their respective KS coordinates $w^\pm$ \eqref{wxcoord} key to the analysis. The ordering of the points has two regimes: firstly, (I)$\cup$(II), i.e.~$t<\frac12(b-a)$,
\EQ{
x^+_{1_L}<x^+_{1_R}<x^+_{2_L}<x^+_{2_R}\ ,\qquad
x^-_{2_L}<x^-_{2_R}<x^-_{1_L}<x^-_{1_R}\ .
\label{sir}
}
In the above, and in the following, we ignore the imaginary shifts in some of the coordinates \eqref{dsd}. The second regime (III)$\cup$(IV)$\cup$(V), $t>\frac12(b-a)$ has,
\EQ{
x^+_{1_L}<x^+_{2_L}<x^+_{1_R}<x^+_{2_R}\ ,\qquad
x^-_{2_L}<x^-_{1_L}<x^-_{2_R}<x^-_{1_R}\ .
\label{sot}
}

It is now simple to write down the entropy of the no-island saddle in the free fermion CFT \cite{Casini:2009sr} in both regimes:
\EQ{
S_\varnothing(A_L\cup A_R)=-\frac c6\sum_{\mu<\nu}(-1)^{\mu-\nu}\log[-(w^+_\mu-w^+_\nu)(w^-_\mu-w^-_\nu)]-\frac c{12}\sum_{\mu}\log|w^+_\mu w^-_\mu|\ ,
\label{pmn}
}
where we label points by their ordering along the Cauchy slice, so $\{2_L,1_L,1_R,2_R\}\to \{1,2,3,4\}$. 
The expression above is the usual CFT result in the thermal state.
The final sum in the above accounts for the conformal factors of the endpoints. To evaluate the entropy at high temperatures ($\beta$ small) it is useful to notice that if $x^\pm_\mu>x^\pm_\nu$, then
\EQ{
\log(w^+_\mu-w^+_\nu) \approx \log w^+_\mu\approx\frac{2\pi}\beta x^+_\mu\ ,\qquad \log(w^-_\nu-w^-_\mu) \approx \log w^-_\nu\approx-\frac{2\pi}\beta x^-_\nu\ ,
}
up to subleading constant pieces. Using these rules, we find that in regime (I)$\cup$(II), with the ordering in \eqref{sir},
\EQ{
S_\varnothing(A_L\cup A_R)=\frac{\pi c}{3\beta}\big(\underbracket{(t+b)+(t+a)}_\text{left}+\underbracket{(t+b)+(t+a)}_\text{right}\big)-\underbracket{\frac{\pi c}{3\beta}(2a+2b)}_\text{conf. factor}=\frac{4\pi ct}{3\beta}\ .
\label{bee}
}
Note that the left-moving and right-moving contributions of all saddles we discuss are equal because of our symmetric choice of $A_L$ and $A_R$. In regime (III)$\cup$(IV)$\cup$(V), with the ordering in \eqref{sot},
\EQ{
S_\varnothing(A_L\cup A_R)=\frac{2\pi c}{3\beta}\big((t+b)+(-t+b)\big)-\frac{\pi c}{3\beta}(2a+2b)=\frac{2\pi c(b-a)}{3\beta}\ .
\label{bea}
}
As expected, the results \eqref{bee} and \eqref{bea} are equal when $t=\frac12(b-a)$ the boundary between regimes (II) and (III).  The no-island saddle corresponds to the BCFT OPE channels (2) and (1), as shown in figure \ref{fig6}.

\subsection{Island saddle}

Now we turn to saddles with islands. In JT gravity the area term in the generalized entropy in \eqref{pii} is given by the value of the dilaton at the putative quantum extremal surface.
We saw in the BCFT analysis that disconnected channels contributing to the entropies arose from one point functions for two out of the four twist operator insertions. This corresponds to an island saddle in JT gravity, arising from two QES, which is, of course, the  minimal number.
There are now six points  along the Cauchy slice $\{2_L,1_L,1_Q,2_Q,1_R,2_R\}$ 
labelled by $\mu=1,2,\ldots,6$ from left to right, with the $\mu=3,4$ labelling the QES. The generalized entropy is explicitly
\EQ{
S_\text{gen}(A_L\cup A_R)&=\underbracket{\frac{\pi c}{6\beta k}\sum_{\mu=1_Q,2_Q}\frac{1-w_\mu^+w_\mu^-}{1+w_\mu^+w_\mu^-}}_\text{dilaton/area}-
\frac c6\sum_{\mu<\nu}(-1)^{\mu-\nu}\log[-(w^+_\mu-w^+_\nu)(w^-_\mu-w^-_\nu)]\\ &-\underbracket{\frac c{12}\sum_{\mu=2_L,1_L,1_R,2_R}\log|w^+_\mu w^-_\mu|}_\text{conf. fact. bath}-\underbracket{\frac c{6}\sum_{\mu=1_Q,2_Q}\log(1+w^+_\mu w^-_\mu)}_\text{conf. fact. AdS}\ ,
\label{pmn2}
}
where the last two sums account for conformal factors from the metric in KS coordinates. The CFT contribution to the entanglement entropy now includes the island interval $I$  between the two QES.

In general, solving the saddle point equations would be a numerical exercise. However, two key simplifications emerge at high temperature limit. Firstly, the QES remain close to the black hole horizon in the semi-classical limit (cf e.g.\cite{Hollowood:2020cou}). Therefore,
assuming that $|w^+_\sigma w^-_\sigma|\ll1$ for $\sigma=1_Q,2_Q$,  the saddle point equations are
\EQ{
w^\mp_\sigma=-s\sum_{\mu(\neq\sigma)}\frac{(-1)^{\sigma-\mu}}{w^\pm_\sigma-w^\pm_\mu}\ .
\label{spp}
}
As previously, we have $s=\beta k/2\pi\ll1$. For simplicity, we will also assume that $\beta|\log s|$, which is the scrambling time of the black hole, is a subleading scale compared with the size of the intervals.

The second key simplification of the high temperature limit yields a picture mirroring the channels of the BCFT calculation in section \ref{sec:bcft} in that the solution can be broken up into the temporal regimes \eqref{nbc} and specifically (II), (III) and (IV). In a given regime, the solutions are dominated by only one term on the right-hand side of \eqref{spp}. Subleading terms affect whether the QES, whilst remaining close to the horizon, are inside or outside of it.

From the analysis of  semi-infinite intervals above, we recall the  appearance of the time scale $t_{\rm exit}=2a$ (assuming small scrambling times) at which the QES exit the horizon.  Whilst there is no discernible signature of this scale in the evolution of the entropy, it proves to be useful  to simplify the discussion by achieving a clean separation of time scales. 

 \paragraph{Regime (IIa):} We will first consider intervals that satisfy the condition $2a < (b-a)/2$, so that $t_{\rm exit}$ lies in the temporal regime (II). Then regime (II) can be split into two sub-regimes:
\be
{\rm IIa}: \quad a<t\leq  2a\,,\qquad {\rm IIb}: \quad 2a< t< \frac12 (b-a)\,.
\ee
The early time period (IIa) is in fact captured by QES locations given by \eqref{QESinf} in the problem with semi-infinite intervals. In this regime, keeping only the dominant terms, the saddle point equations are (assuming $|w_{1_Q}^-|\gg1$, $|w_{1_Q}^+w_{1_Q}^-|\ll1$)
\be
w^-_{1_Q}\approx\frac{s}{w^+_{1_Q}-w^+_{1_L}}\,,\qquad\qquad 
\frac1s w^+_{1_Q}\approx-\frac{1}{w_{1_L}^-}+\frac{1}{w_{1_Q}^-}-\frac{1}{w^-_{1_Q}-w_{1_R}^-}\,,
\ee
with corresponding conditions for the second QES. In the symmetric situation $w_{1_Q}^\pm=w_{2_Q}^\mp$. These conditions yield the locations  \eqref{QESinf}. Explicitly,
\be
 w_{1_Q}^+= -se^{-2\pi(t+a)/\beta}+\frac{1}{s}e^{-6\pi(t-a)/\beta}\,,\quad w_{1_Q}^-=se^{2\pi(t-a)/\beta}\,.
\ee

In each of the remaining temporal regimes, the solutions are dominated by only one term on the right-hand side of \eqref{spp} say $\mu=\mu(\sigma)$:
\EQ{
w^\mp_\sigma\approx\frac s{w^\pm_\sigma-w^\pm_{\mu(\sigma)}}\ .
\label{bbc}
}
In the above, we have assumed that $(-1)^{\sigma-\mu(\sigma)}=-1$ which is true for all the examples. 
The terms that have been neglected are either subleading or cancel out in pairs.

The simpler equations \eqref{bbc} are quadratic with only one of the solutions physically consistent, i.e.~has $|w^+_\sigma w^-_\sigma|\ll1$ and the QES being points on a Cauchy slice containing the points in the baths:
\EQ{
w^\pm_\sigma=\frac{w^\pm_{\mu(\sigma)}}2\Big(1-\sqrt{1+\frac{4s}{w^+_{\mu(\sigma)}w^-_{\mu(\sigma)}}}\Big)\ .
\label{jjn}
}
In the high temperature limit, this has a crossover between two distinct regimes. Firstly, when $x_{\mu(\sigma)}^+>x^-_{\mu(\sigma)}$, in which case the solution becomes
\EQ{
w^\pm_\sigma\approx-\frac s{w^\mp_{\mu(\sigma)}}\,.
\label{pep}
}
According to \eqref{pep}, in this regime, a QES corresponds to the following pattern of coordinates:
\EQ{
\begin{tikzpicture}
\node[right] at (0,0) {$\cdots<x^+_\sigma~<~x^+_{\mu(\sigma)}<\cdots$};
\node[rotate=90] at (1.4,-0.6) {$=$};
\node[rotate=90] at (2.8,-0.6) {$=$};
\node[right] at (0,-1.2) {$\cdots<x_{\mu(\sigma)}^-<\,x_\sigma^-<\cdots$};
\end{tikzpicture}
\label{pat}
}
For these solutions it follows from \eqref{pep} that the allowed values for $\pm x^\pm_\sigma$ are from the set $\{-t-b,-t-a,t-b,t-a\}$.  

The solution \eqref{jjn} then has a crossover to the regime where $x_{\mu(\sigma)}^+<x^-_{\mu(\sigma)}$ for which
\EQ{
w^\pm_\sigma\approx\sqrt{s\frac{w^\pm_{\mu(\sigma)}}{w^\mp_{\mu(\sigma)}}}\ .
\label{pep2}
}
In this case, the pattern is 
\EQ{
\begin{tikzpicture}
\node[right] at (0,0) {$\cdots<x^+_{\mu(\sigma)}<x^+_\sigma<\cdots$}; 
\node[right] at (1.4,-1.2) {$\cdots<x^-_\sigma<x_{\mu(\sigma)}^-<\cdots$};
\node at (2.8,-2.4) {$\tfrac12(x^+_{\mu(\sigma)}+x^-_{\mu(\sigma)})$};
\node[rotate=90] at (2.8,-0.6) {$=$};
\node[rotate=90] at (2.8,-1.8) {$=$};
\end{tikzpicture}
\label{pat2}
}
 Again, we assume large intervals $b>3a$ and use the analysis above to piece together the island saddle.
 
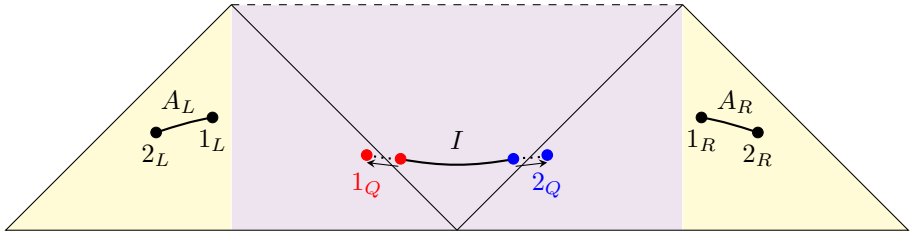
\begin{figure}[ht]
\begin{center}
\begin{tikzpicture}[scale=1]
\draw[fill=yellow!20,yellow!20] (-6,0) -- (-3,0) -- (-3,3) -- cycle;
\draw[fill=yellow!20,yellow!20] (6,0) -- (3,0) -- (3,3) -- cycle;
\draw[white,fill=Plum!10!white] (-3,0) -- (3,0) -- (3,3) -- (-3,3) -- cycle;
\draw[-] (-3,3) -- (-6,0) -- (6,0) -- (3,3);
\draw[dashed] (-3,3) -- (3,3);
\draw[-] (0,0) -- (3,3);
\draw[-] (-3,3) -- (0,0);
\filldraw[black] (3.25,1.5) circle (2pt);
\filldraw[black] (-3.25,1.5) circle (2pt);
\filldraw[black] (4,1.3) circle (2pt);
\filldraw[black] (-4,1.3) circle (2pt);
\node at (-4,1) {\footnotesize$2_L$};
\node at (-3.25,1.2) {\footnotesize $1_L$};
\node[red] at (-1.2,0.6) {\footnotesize $1_Q$};
\node[blue] at (1.2,0.6) {\footnotesize $2_Q$};
\node at (3.25,1.2) {\footnotesize $1_R$};
\node at (4,1) {\footnotesize $2_R$};
\draw[thick] (-3.25,1.5) to[out=-170,in=18] (-4,1.3);
\draw[thick] (3.25,1.5) to[out=-10,in=162] (4,1.3);
\draw[thick] (-0.78,0.95) to[out=-10,in=-170] (0.78,0.95);
\draw[dotted,thick] (-1.2,1) to[out=-10,in=-180] (-0.78,0.95);
\draw[dotted, thick] (0.78,0.95) to[out=10,in=-160] (1.2,1);
\draw [-stealth](0.78,0.85) -- (1.2,0.9);
\draw [-stealth](-0.78,0.85) -- (-1.2,0.9);
\filldraw[blue] (1.2,1) circle (2pt);
\filldraw[blue] (0.75,0.95) circle (2pt);
\filldraw[red] (-1.2,1) circle (2pt);
\filldraw[red] (-0.75,0.95) circle (2pt);
\node at (0,1.2) {\footnotesize $I$};
\node at (3.7,1.7) {\footnotesize $A_R$};
\node at (-3.7,1.7) {\footnotesize $A_L$};
\end{tikzpicture}
\caption{\footnotesize The island saddle in regime (II) has an island with QES that can move from inside to outside the horizon (when $2a < \frac12(b-a)$) as shown in red and blue. If $2a > \frac12(b-a)$, they remain inside the horizon. The labelling of points along a Cauchy slice are shown and only the upper half of the Penrose diagram in figure \ref{fig1} is shown.}
\label{fig10} 
\end{center}
\end{figure}

\paragraph{Regime (IIb):} In regime (IIb) which matches smoothly with (IIa), we find that the island saddle is of type \eqref{pep} with 
\EQ{
w^\pm_{1_Q}=-\frac s{w^\mp_{1_L}}=\mp se^{2\pi(\mp t-a)/\beta}\ ,\quad w^\pm_{2_Q}=-\frac s{w^\mp_{1_R}}=\pm se^{2\pi(\pm t-a)/\beta}\ ,
\label{bxx}
}
which are outside the horizon on the left and right (since $w_3^+,w_4^-<0$), respectively, as shown in figure \ref{fig10}. The orderings of the $x^\pm$ coordinates are
\EQ{
{\color{red}x^+_{1_Q}}<x^+_{1_L}<{\color{blue}x^+_{2_Q}}<x^+_{1_R}<x^+_{2_L}<x^+_{2_R}\ ,\quad
x^-_{2_L}<x^-_{2_R}<x^-_{1_L}<{\color{red}x^-_{1_Q}}<x^-_{1_R}<{\color{blue}x^-_{2_Q}}\ ,
\label{soy}
}
where the QES coordinates are shown in red and blue.  One may check that the conditions are satisfied for this to be a solution of the saddle point equations. In particular, 
\EQ{
|w^+_{1_Q}w^-_{1_Q}|=|w^+_{2_Q}w^-_{2_Q}|=s^2e^{-4\pi a/\beta}\ll1
}
and the QES come in the pattern \eqref{pat}.

The entropy is given in \eqref{pmn2} making the approximation $|w^+_\sigma w^-_\sigma|\ll1$ for $\mu=1_Q,2_Q$, so that the dilaton or area term becomes a constant and the conformal factor term in the AdS region---the last sum---can be neglected. We find
\EQ{
S_I(A_L\cup A_R)&=\frac{\pi c}{3\beta k}+\frac{2\pi c}{3\beta}\big((t+b)+(t+a)+(-t+a)\big)-\frac{\pi c}{3\beta}(2a+2b)\\ &=\frac{2\pi c}{3\beta}\big(\frac1{2k}+t+a\big)\ .
\label{bed}
}
The constant term, proportional to $1/k$, is twice the Bekenstein-Hawking entropy of the black hole. Comparing with the BCFT result \eqref{BCFTentropy}, this is exactly the contribution from the OPE channel (3) which dominates in region (II). Note that the entropy in sub-regimes (IIa)  and (IIb) are given by the same expression up to  exponentially small corrections. We can, in fact make an identification between the boundary entropy and  black hole entropy, even though JT gravity does not describe the BCFT limit in the holographic dual,
\be
\log g_b \quad\longleftrightarrow\quad S_\text{BH}^{(\beta)}=\frac{\pi c}{6\beta k}\ .
\ee
We view the black hole entropy as a temperature dependent boundary entropy, and the free fermion BCFT limit wherein $\log g_b =0$ is like an infrared limit in which typical length and time scales are taken to be much larger than $1/k$
(in conjunction with the high temperature thermodynamic limit).

Within the island saddle in regime (II) we can identify how the {\em bulk} OPE channel which dominates the entropy matches  the OPE channel in the BCFT description. In this regime, the motion and the precise location of the exiting QES does not affect the leading behaviour of the entropy  which is determined by the bulk OPE channel $w^\pm_{1_L}\to w^\pm_{1_Q}$,  $w^\pm_{1_R}\to w^\pm_{2_Q}$ and $w^\pm_{2_L}\to w^\pm_{2_R}$:
\bea
S_{I}(A_L\cup A_R)&&=2 S_{\rm BH}+\frac{c}{6}\left(\log{\sigma_{1_L,1_Q}+\log\sigma_{2_Q,1_R}+\log\sigma_{2_L,2_R}}\right)\\\nonumber
&&-\frac{c}{12}\sum_{\mu=1_{L,R},2_{L,R}}\log |w_\mu^+ w_\mu^-|\,,
\eea
yielding the result \eqref{bed}, where $\sigma_{\mu,\nu}=-(w_\mu^+-w_\nu^+)(w_\mu^--w_\nu^-)$. The result tells us that contractions of  twist and anti-twist operator  insertions at a bulk interval endpoint and  the QES,  act to reproduce contractions between points and images in the BCFT picture. 
This yields the disconnected portion of the BCFT OPE channel (3) (see figure \ref{fig4}), and reveals how the correspondence between  BCFT image points and QES surfaces should work.

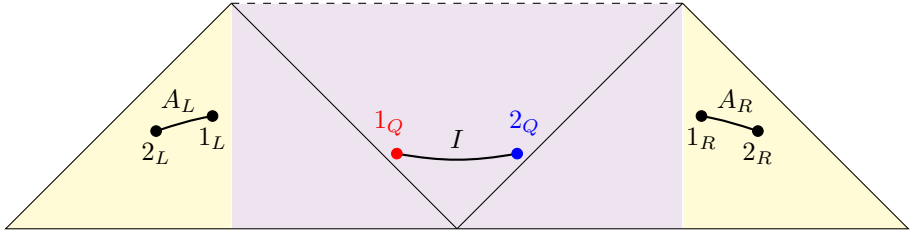
\begin{figure}[ht]
\begin{center}
\begin{tikzpicture}[scale=1]
\draw[fill=yellow!20,yellow!20] (-6,0) -- (-3,0) -- (-3,3) -- cycle;
\draw[fill=yellow!20,yellow!20] (6,0) -- (3,0) -- (3,3) -- cycle;
\draw[white,fill=Plum!10!white] (-3,0) -- (3,0) -- (3,3) -- (-3,3) -- cycle;
\draw[-] (-3,3) -- (-6,0) -- (6,0) -- (3,3);
\draw[dashed] (-3,3) -- (3,3);
\draw[-] (0,0) -- (3,3);
\draw[-] (-3,3) -- (0,0);
\filldraw[black] (3.25,1.5) circle (2pt);
\filldraw[black] (-3.25,1.5) circle (2pt);
\filldraw[black] (4,1.3) circle (2pt);
\filldraw[black] (-4,1.3) circle (2pt);
\node at (-4,1) {\footnotesize $2_L$};
\node at (-3.25,1.2) {\footnotesize $1_L$};
\node[red] at (-0.9,1.4) {\footnotesize $1_Q$};
\node[blue] at (0.9,1.4) {\footnotesize $2_Q$};
\node at (3.25,1.2) {\footnotesize $1_R$};
\node at (4,1) {\footnotesize $2_R$};
\draw[thick] (-3.25,1.5) to[out=-170,in=18] (-4,1.3);
\draw[thick] (3.25,1.5) to[out=-10,in=162] (4,1.3);
\draw[thick] (-0.8,1) to[out=-10,in=-170] (0.8,1);
\filldraw[blue] (0.8,1) circle (2pt);
\filldraw[red] (-0.8,1) circle (2pt);
\node at (0,1.2) {\footnotesize $I$};
\node at (3.7,1.7) {\footnotesize $A_R$};
\node at (-3.7,1.7) {\footnotesize $A_L$};
\end{tikzpicture}
\caption{\footnotesize The island saddle in regime (III)$\cup$(IV) has an island with QES behind the horizon as shown in red and blue.}
\label{fig11} 
\end{center}
\end{figure}

\paragraph{Regime (III):} Now we transition into regime (III) for which the island saddle is of type \eqref{pep} with QES at
\EQ{
w^-_{1_Q}=-\frac s{w^+_{1_L}}=se^{2\pi(t-a)/\beta}\ ,\qquad w^+_{1_Q}=-\frac s{w^-_{2_R}}=se^{2\pi(t-b)/\beta}\ ,\\
w^-_{2_Q}=-\frac s{w^+_{2_L}}=se^{2\pi(t-b)/\beta}\ ,\qquad w^+_{2_Q}=-\frac s{w^-_{1_R}}=se^{2\pi(t-a)/\beta}\ ,
}
which are inside the horizon as shown in figure \ref{fig11}. 
So as $t$ increases from (II) to (III) the QES move from outside to inside the horizon if $2a < \tfrac12(b-a)$, or simply continue to stay inside if $2a > \tfrac12(b-a)$. Later we will show how this happens precisely. The orderings of the $x^\pm$ coordinates  in this regime are,
\EQ{
{\color{red}x^+_{1_Q}}<x^+_{1_L}<{\color{blue}x^+_{2_Q}}<x^+_{2_L}<x^+_{1_R}<x^+_{2_R}\ ,\quad
x^-_{2_L}<x^-_{1_L}<x^-_{2_R}< {\color{red}x^-_{1_Q}}<x^-_{1_R}<{\color{blue}x^-_{2_Q}}
\label{swy2}
}
and
\EQ{
|w^+_{1_Q}w^-_{1_Q}|=|w^+_{2_Q}w^-_{2_Q}|=s^2e^{2\pi(2t-a-b)/\beta}\ll1\ ,
}
since $t<\frac12(a+b)$ in (III). In addition, the other consistency conditions are satisfied and the QES come in the pattern \eqref{pat}. 

The entropy in regime (III) is
\EQ{
S_I(A_L\cup A_R)&=\frac{\pi c}{3\beta k}+\frac{2\pi c}{3\beta}\big((t+b)+(-t+b)+(-t+a)\big)-\frac{\pi c}{3\beta}(2a+2b)\\ &=\frac{2\pi c}{3\beta}\big(\frac1{2k}-t+b\big)\ ,
\label{beq}
}
matching with the BCFT result \eqref{BCFTentropy}.  The agreement here is important, as it confirms the existence of the dip in the entropies. In this regime, the BCFT channel (4) dominates the entropy. The channel is characterized by contraction of a pair of twist-anti-twist operators in  CFT$_R$, with BCFT images in CFT$_L$ (figure \ref{fig4}).  In fact, twist insertions at the near endpoint $x=a$ in one CFT are contracted with images at the far endpoint $x=b$ in the other CFT. As remarked earlier, these are lightcone singularities with a natural interpretation in free CFTs or those with a quasiparticle interpretation. In the bulk CFT, the dominant terms in regime (III) arise from the contractions $w_{1_L}^+\to w_{1_Q}^+$, $w_{1_Q}^-\to w_{2_R}^-$, $w_{2_L}^+\to w_{2_Q}^+$, $w_{2_Q}^-\to w_{1_R}^-$, $w_{2_L}^-\to w_{1_L}^-$,  $w_{1_R}^+\to w_{2_R}^+$:
\bea
S_I(A_L\cup A_R)&&=2S_{\rm BH}+\frac{c}{6}\log\left|(w_{1_L}^+ - w_{1_Q}^+)(w_{1_Q}^- - w_{2_R}^-)(w_{2_L}^+- w_{2_Q}^+)(w_{2_Q}^-- w_{1_R}^-)\right|+\nonumber\\\nonumber\\
&&\frac{c}{6}\log\left|(w_{2_L}^- - w_{1_L}^-)(w_{1_R}^+ - w_{2_R}^+)\right|  -\frac{c}{12}\sum_{\mu=1_{L,R},2_{L,R}}\log |w_\mu^+ w_\mu^-|\,.\label{bulkiii}
\eea
Note now that the  effect of the singularity $\sim \log\sinh\frac\pi\beta(x_{2_L}^+-x_{1_R}^-)$ in the BCFT channel (4)
  is reproduced by the successive bulk contractions $\sim \log (w_{2_L}^+- w_{2_Q}^+)(w_{2_Q}^-- w_{1_R}^-)$ via an intermediate QES.
Evaluating this (in the high temperature limit) gives us the result for regime (III) as before, but importantly it explicitly demonstrates the role played by the QES in reproducing BCFT channel (4).

\paragraph{Regime (IV):} Finally the island saddle of (III) transitions into (IV) as a solution of type \eqref{pep2},
\EQ{
w_{1_Q}^\pm=\sqrt{s}e^{\mp\pi(b-a)/\beta}\ ,\qquad w^\pm_{2_Q}=\sqrt{s}e^{\pm \pi(b-a)/\beta}\ ,
}
which remain inside the horizon. In this case the coordinates satisfy the consistency conditions for the approximations as long as $s$ is small,
\EQ{
|w^+_{1_Q}w^-_{1_Q}|=|w^+_{2_Q}w^-_{2_Q}|=s\ll1\,.
} 
The ordering of the $x^\pm$ coordinates is
\EQ{
x^+_{1_L}<{\color{red}x^+_{1_Q}}<x^+_{2_L}<{\color{blue}x^+_{2_Q}}<x^+_{1_R}<x^+_{2_R}\ ,\quad
x^-_{2_L},<x^-_{1_L}<{\color{red}x^-_{1_Q}}<x^-_{2_R}<{\color{blue}x^-_{2_Q}}<x^-_{1_R}
\label{swy3}
}
and the QES come in the pattern \eqref{pat2}. The entropy is
\EQ{
S_I(A_L\cup A_R)&=\frac{\pi c}{3\beta k}+\frac{2\pi c}{3\beta}\big((t+b)+\tfrac12(b-a)+\tfrac12(a-b)\big)-\frac{\pi c}{6\beta}(4a+4b)\\ &=\frac{2\pi c}{3\beta}\big(\frac1{2k}+t-a\big)\,,
\label{beq2}
}
matching the rising portion of the BCFT channel (4) contribution, following the dip in regime (III). As in the case of the BCFT, both the fall and rise are governed by the same bulk channel \eqref{bulkiii}. The actual magnitudes of the coordinates of the QES do not contribute at leading order to the final result for the high temperature entropy, but how big they are relative to coordinates of other points determines whether we are in regime (III) or (IV).

The final entropy is obtained by taking the minimum of the island and no-island saddles. This gives the result shown in figure \ref{fig6} plotted in the case where the black hole entropy constant is small. The no-island saddle is shown as the black line and the island saddle as the red line.

\pgfdeclareimage[interpolate=true,width=7cm]{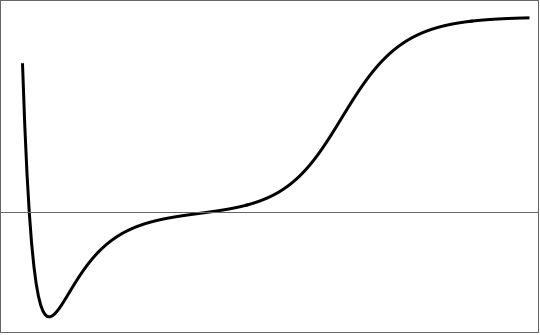}{picnew}
\pgfdeclareimage[interpolate=true,width=7cm]{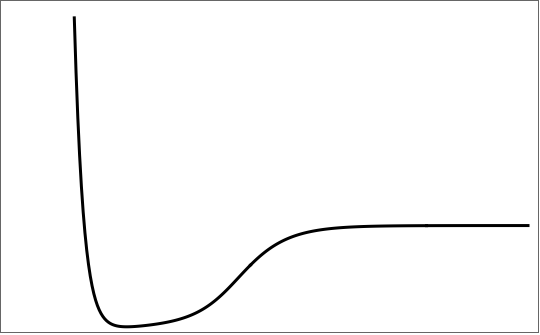}{picnew1}
\begin{figure}[h]
\begin{center}
\begin{tikzpicture}[scale=1]
\pgftext[at=\pgfpoint{-6.9cm}{0cm},left,base]{\pgfuseimage{picnew}} 
\pgftext[at=\pgfpoint{0.7cm}{0cm},left,base]{\pgfuseimage{picnew1}}
\node at (-6.5,-0.17) {\footnotesize $2a$};
\node at (7.6,-0.1) {\footnotesize $b$};
\node at (-0.1,-0.1) {\footnotesize $b$};
\node at (-4.3,-0.23) {\footnotesize $\tfrac{b-a}2$};
\node at (-2.5,-0.23) {\footnotesize $\tfrac{a+b}2$};
\node at (2.2,-0.23) {\footnotesize $\tfrac{b-a}2$};
\node at (3.9,-0.23) {\footnotesize $\tfrac{a+b}2$};
\node at (4.5,-0.7) {\footnotesize $t$};
\draw [-stealth](4.6,-0.7) -- (5.1,-0.7);
\node at (-4,-0.7) {\footnotesize $t$};
\draw [-stealth](-3.9,-0.7) -- (-3.4,-0.7);
\draw[dotted] (-6.55,0.05) -- (-6.55,1.5);
\draw[dotted,thick] (-4.3,0.05) -- (-4.3,4.3);
\draw[dotted,thick] (-2.5,0.05) -- (-2.5,4.3);
\draw[dotted,thick] (2.2,0.05) -- (2.2,4.3);
\draw[dotted,thick] (3.9,0.05) -- (3.9,4.3);
\node at (-7.4,2.85) {\footnotesize $w^-_{2_Q}$};
\draw [-stealth](-7.0, 2.3) -- (-7.0,3.1);
%\node at (0.5,2.85) {\footnotesize $w^-_{2_Q}$};
\node at (-7.1,1.6) {\footnotesize 0};
\node at (0.5,0.1) {\footnotesize 0};
\node at (1.4,2.9) {\footnotesize (II)};
\node at (-5.6,2.9) {\footnotesize (II)};
\node at (-3.4,2.9) {\footnotesize (III)};
\node at (3.1,2.9) {\footnotesize (III)};
\node at (-1.2,2.9) {\footnotesize (IV)};
\node at (5.8,2.9) {\footnotesize (IV)};
\node at (-1,1.6) {\footnotesize horizon};
\node at (6,0.1) {\footnotesize horizon};
\node at (6,0.6) {\footnotesize inside};
\node at (-1,2.2) {\footnotesize inside};
\node at (-1,1) {\footnotesize outside};
\end{tikzpicture}
\caption{\footnotesize The $w^-_{2_Q}$ coordinate of the right QES determined numerically with some large but finite temperature. When $2a < \frac12(b-a)$ (left panel), the QES starts off inside the horizon, exits at $t=2a$ at the onset of regime (II) and plunges back in at $t=\frac12(b-a)$ to remain inside in regime (III), before saturating to the constant in region (IV). The situation when $2a>\frac12(b-a)$ (right panel) shows the corresponding motion of the QES behind the horizon at all times.}
\label{fig12}
\end{center}
\end{figure}
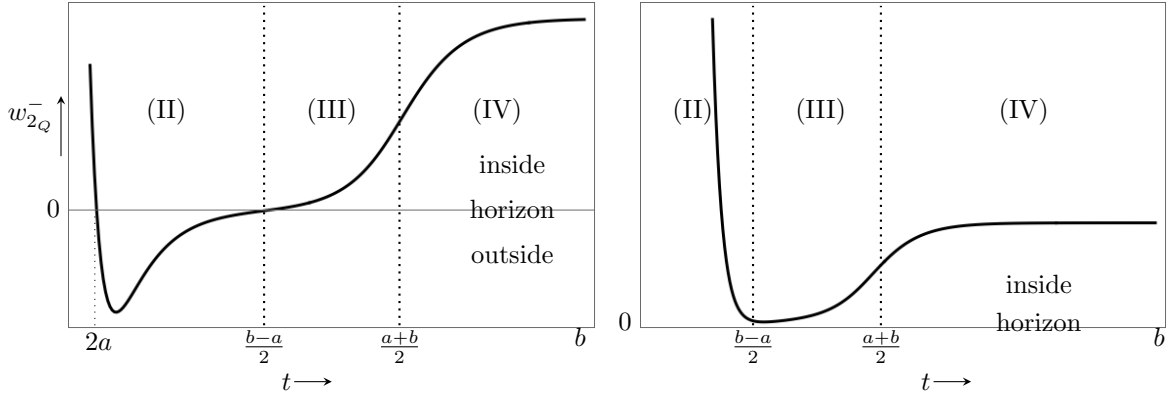

The regimes (IIa) and (IIb) show no qualitative difference insofar as the entropy evolution is concerned. When $2a> (b-a)/2$, the two sub-regimes merge into one and the QES stays inside the horizon at all times. Thus regime (IIb) is eliminated in this case while the  analysis of regimes (III) and (IV) remains largely unchanged, except for the transient position of the QES in regime (III) which receives a correction term whilst remaining inside the horizon.

What our approximation to the saddle point equations in \eqref{bbc} obscures is that the island saddles that dominate the entropy in the regimes (II), (III) and (IV) are in fact smoothly related across the boundaries in a way that becomes sharper as the temperature increases. We can do better by working with a refinement of the approximation \eqref{bbc} which includes all the relevant terms for describing the island saddle across (II), (III) and (IV): 
\EQ{
w_{1_Q}^- &= \frac{s}{w_{1_Q}^+ - w_{1_L}^+}\,, \qquad w_{1_Q}^+ =\frac{s}{w_{1_Q}^-}-\frac{s}{w_{1_Q}^--w_{1_R}^-}+\frac{s}{w_{1_Q}^- - w_{2_R}^-} - \frac{s}{w_{1_L}^-} \ ,\\[5pt]
w^+_{2_Q}&=\frac s{w^-_{2_Q}-w^-_{1_R}}\ ,\qquad w^-_{2_Q}=\frac{s}{w_{2_Q}^+}-\frac{s}{w_{2_Q}^+-w_{1_L}^+}+\frac s{w^+_{2_Q}-w^+_{2_L}}-\frac s{w^+_{1_R}}\ .
\label{huy}
}
In the last terms, we have used the fact that $|w^-_{1_Q}|\ll |w^-_{1_L}|$, $|w^+_{2_Q}|\ll |w^+_{1_R}|$, $|w^+_{1_Q}|\ll|w^-_{1_Q}|$, and  $|w^-_{2_Q}|\ll|w^+_{2_Q}|$ respectively. Remarkably, each pair of equations when combined results in  just a quadratic equation  for $w_{1Q}^+$ and $w_{2Q}^-$: 
\bea
&&\frac{1}{w_{1_Q}^- - w_{2_R}^-}-\frac{1}{w_{1_Q}^- - w_{1_R}^-} -\left(\frac{1}{w_{1_L}^-}+\frac1s w_{1_L}^+\right)=0\,,\nonumber\\\\\nonumber
&&\frac{1}{w_{2_Q}^+ - w_{2_L}^+}-\frac{1}{w_{2_Q}^+ - w_{1_L}^+} -\left(\frac{1}{w_{1_R}^+}+\frac1s w_{1_R}^-\right)=0\,,
\eea

Solutions to these quadratic equations smoothly interpolate the asymptotic solutions in regimes (II), (III) and (IV). Taking the right QES, for small $\beta$ we can write the solution for the $w^-$ coordinate (after neglecting terms that are consistently small),
\bea
w^-_{2_Q}= &&2s\left(\sqrt{4s^2 e^{2\pi(b-3a)/\beta}+4s e^{2\pi(b-a)/\beta}+e^{4\pi(b-t)/\beta}-2e^{2\pi(b+a-2t)/\beta}}\right.\label{qss}\\\nonumber
&&\left.-e^{2\pi(a-t)/\beta}-e^{2\pi(b-t)/\beta}\right)^{-1}\,-\, e^{-2\pi(t-a)/\beta}\,.
%\frac{e^{2\pi(a_2-t)/\beta}}2\Big(\sqrt{1-2se^{-4\pi a_2/\beta}+4s e^{2\pi(2t-a_2-b_1)/\beta}}-1-se^{-4\pi a_2/\beta}\Big)\ .
\eea
%\EQ{
%w^-_{2_Q}=\frac{e^{2\pi(a_2-t)/\beta}}2\Big(\sqrt{1-se^{-4\pi a_2/\beta}+4s e^{2\pi(2t-a_2-b_1)/\beta}}-1\Big)\ .
%\label{qss}
%}
%We have written the result for the general non-symmetric case for later use. There is a similar expression for $w^+_{1_Q}$ with $a_2\to a_1$ and $b_1\to b_2$. 
The expression \eqref{qss} displays the three regimes (II)$\to$(III)$\to$(IV) with transitions at $t=\frac12(b\pm a)$ as $t$ increases. In particular, for the situation with $2a < \frac12(b-a)$, The QES moves out of the horizon at $t=2a$ and as (II) crosses over to (III) at $t=\tfrac12(b-a)$, the QES moves smoothly from outside to inside the horizon as shown in figure \ref{fig12}. The left QES behaves in a similar way.

\subsection{Small intervals}

When $b<3a$ there is a different pattern of saddles 
\EQ{
\text{(I)}&=\{0< t< \tfrac12(b-a)\}\ ,\qquad~\, \text{(V)}=\{\tfrac12(b-a)< t<a\}\ ,\\[5pt] \text{(III)}&=\{a< t<\tfrac12(a+b)\}\ ,\qquad \text{(IV)}=\{\tfrac12(a+b)< t< b\}\ ,\\[5pt] \text{(V)}&=\{b< t\}\ .
\label{nbc2}
}
The labelling here matches the labelling of the saddles in the last section. So in this case, the system reaches equilibrium early at $t=\tfrac12(b-a)$ but then there is a dip later on. In this case, the QES are always inside the horizon in the island saddle since regime (II) does not occur. The entropy is shown in figure \ref{fig7} and, again, matches the BCFT result perfectly.

\subsection{Generic intervals}

We now consider the case with two generic intervals. This brings additional regimes controlled by different BCFT channels into play, as shown by the facet-like structure in figure \ref{fig9}. In particular, the crossover between regimes (II) and (III) gets replaced by a new regime  dominated by the disconnected BCFT  channel of type (5) shown in figure \ref{fig8}, or of the same type with $L$ and $R$ exchanged. Based on  our observations in the symmetric case, we expect channel (5) to correspond to the generalized entropy in the presence of 2 QES dominated by the bulk channels: $w_{1_L}^+\to w_{1_Q}^+$, $w_{1_Q}^-\to w_{2_R}^-$, $w_{1_R}^+\to w_{2_Q}^+$, $w_{2_Q}^-\to w_{1_R}^-$, $w_{2_L}^+\to w_{2_R}^+$, $w_{2_L}^-\to w_{1_L}^-$. A quick calculation confirms this as the correct interpretation, assuming the QES are close to the horizon and their coordinates are relatively small.

The equations for the QES are still \eqref{huy} and the solution for the right QES $w^-_{2_Q}$ coordinate can be obtained  analogously to \eqref{qss} . It is then apparent that the left and right QES behave independently. Taking the situation with the exit times $t_{\rm exit}$\footnote{The  left/right QES exit the horizon at $t_{\rm exit}= \frac12(3 a_{1,2} + a_{2,1})$ (see eq.\eqref{QESinf}), provided these are small compared with subsequent transition time scales below.} sufficiently small,  the left/right QES has three regimes with transitions at $t=\frac12(b_2\pm a_1)$ and $t=\frac12(b_1\pm a_2)$, respectively. In particular, the QES plunges into the horizon at the lower transition and since this is different for the left and right QES there will be configurations where one QES is outside and the other is inside the horizon. For the example in figure \ref{fig9}, the right QES plunges first at $t=\frac12(b_1-a_2)=4.5$ as we expected above. This is the regime labelled (IIA) where the entropy plateaus. The left QES then plunges at $t=\frac12(b_2-a_1)=6.65$ as (IIA) transitions to (III) and the entropy starts to decrease into the dip. The dip itself is also smoothed off in regime (IIIA), since the final transition occurs at different times, $t=\frac12(a_2+b_1)=8$ for the right QES and $t=\frac12(a_1+b_2)=8.65$ for the left QES.

\section{Geodesic approximation}
\label{s5}

At high temperature the entanglement structure of the state can be visualized in terms of the positions of localized wave packets, since modes of characteristic energy $\beta^{-1}$ can be localized on distance scales $\sim\beta$ which is  small as $\beta\to0$. A left-moving\footnote{We use the convention that a left-moving mode means with respect to our figure, so on CFT$_L$ with $x$ increasing and on CFT$_R$ with $x$ decreasing, or in the black hole outgoing on the left and infalling on the right.} wave packet on the left localized around a geodesic, or ray, at fixed  $x^+=\lambda$ is entangled with a localized wave packet on the right with the reflected coordinates $x^+=\lambda$ (and similarly for the right-moving modes). In this limit, we can analyse the state and the entropy of $A_L\cup A_R$ by a simple process of ray tracing.

\subsection{BCFT}

\begin{figure}[ht]
\begin{center}
\begin{tikzpicture}[scale=0.7]
\draw[white,fill=red!30!white,opacity=0.4] (1,4.75) -- (0,3.75) -- (0,0) -- (3.75,0) --  (1.875,1.875) -- (4.75,4.75) -- cycle;
\draw[white,fill=Emerald!30!white,opacity=0.4] (-6,4.75) -- (-1.25,0) -- (0,0) -- (-4.75,4.75) -- cycle;
\draw[white,fill=black!30!white,opacity=0.4] (5.75,0) -- (1,4.75) -- (6,4.75) -- (8,2.75) to[out=-80,in=100] (8,0) -- cycle;
\draw[black!30,<->,thick] (-7.5,0) -- (9,0);
\draw[black!30,->,thick] (0,0) -- (0,6.5);
\draw[thick,Emerald!50] (0,0) -- (-4.75,4.75);
\draw[thick,Emerald!50] (-6,4.75) -- (-1.25,0);
\draw[black!50,thick] (1,4.75) -- (5.75,0);
\draw[black!50,thick] (6,4.75) -- (8,2.75);
\draw[red!50,thick] (1,4.75) -- (0,3.75) -- (3.75,0);
\draw[red!50,thick] (0,0) -- (4.75,4.75);
\node at (0,7.1) {$t$};
\node at (9.5,0) {$x$};
\node at (-8,0) {$x$};
%\node[red] at (0,3.5) {$\tilde I^{(r)}$};
\node at (6,2.5) {$A_R^{(l)}$};
\node at (3.5,5.2) {$A_R$};
\node at (-3.5,5.2) {$A_L$};
\node[red] at (1.8,0.8) {$\hat A_R^{(l)}$};
\node[Emerald] at (-3,2.3) {$A_L^{(l)}$};
\draw[very thick] (6,4.75) -- (1,4.75);
\filldraw (6,4.75) circle (2pt);
\filldraw (1,4.75) circle (2pt);
\draw[very thick] (-6,4.75) -- (-1,4.75);
\filldraw (-6,4.75) circle (2pt);
\filldraw (-1,4.75) circle (2pt);
\end{tikzpicture}
\caption{\footnotesize The left-moving modes that pass through $A_L\cup A_R$ at time $t$ are split into the three subsets as indicated. Here, $t$ is in the temporal interval (IV).}
\label{fig13}
\end{center}
\end{figure}
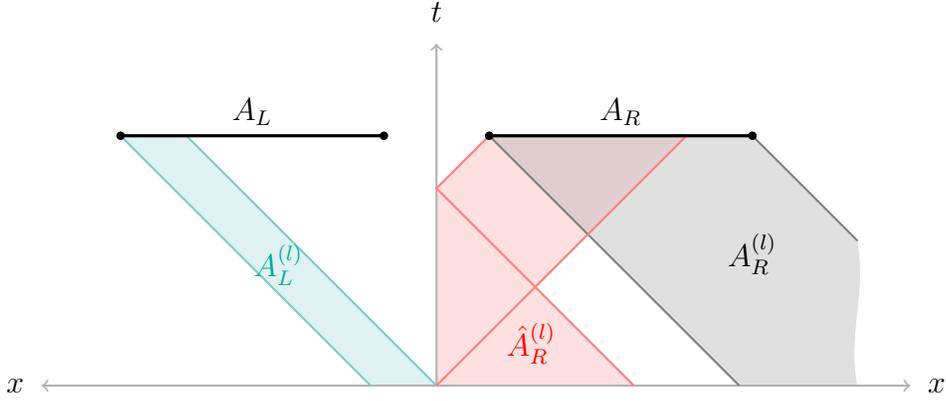

We start with the BCFT. Referring to figure \ref{fig13}, consider left-moving modes that pass through $A_R$ at time $t$
 (there is a similar story for the right-moving modes). There are two distinct set of modes: firstly modes in the interval $x^+\in A_R^{(l)}\equiv[a+t,b+t]$ that pass through $A_R$ as left-moving modes. But there are left-moving modes in interval $x^+\in\hat A_R^{(l)}\equiv[\max(t-b,0),\max(t-a,0)]$ that reflect off the boundary and pass through $A_R$ as right-moving modes. On the left, there are no reflected modes to consider and the left-moving modes that pass through $A_L$ at time $t$ are $x^+\in A_L^{(l)}\equiv[\max(a-t,0),\max(b-t,0)]$. 
 The three subsets of modes are shown in figure \ref{fig13}.

\begin{figure}[ht]
\begin{center}
\begin{tikzpicture}[scale=0.7]
\draw[white,fill=red!30!white,opacity=0.2] (0,1) -- (2.5,3.5) -- (0,6) -- cycle;
\draw[white,fill=Emerald!30!white,opacity=0.2] (1,0) -- (6,0) -- (3.5,2.5) -- cycle;

\draw[white,fill=Emerald!30!white,opacity=0.2] (1,0) -- (6,0) -- (3.5,2.5) -- cycle;
\draw[white,fill=black!30!white,opacity=0.2] (6,0) -- (1,0) -- (8,7) to[out=-80,in=100] (8,2) -- cycle;
\draw[blue!20,fill=blue!20] (1,0) -- (0,1) -- (2.5,3.5) -- (3.5,2.5) -- cycle;
\draw[blue!20,fill=blue!20] (6,0) -- (3.5,2.5) -- (8,7) to[out=-80,in=100] (8,2) -- cycle;
\draw[blue!20,fill=blue!20] (2.5,3.5) -- (6,7) to[out=-170,in=10] (1,7) -- (0,6) -- cycle;
\draw[black!30,->,thick] (0,0) -- (9,0);
\draw[black!30,->,thick] (0,0) -- (0,7.5);
\draw[line width=1mm,Emerald] (1,0) -- (0,1);
\draw[line width=1mm,Emerald] (6,0) -- (0,6);
\draw[line width=1mm] (1,0) -- (8,7);
\draw[line width=1mm] (6,0) -- (8,2);
\draw[red,line width=1mm] (0,1) -- (6,7);
\draw[red,line width=1mm] (0,6) -- (1,7);
\draw[dotted] (-1,1) -- (8,1);
\draw[dotted] (-1,2.5) -- (8,2.5);
\draw[dotted] (-1,3.5) -- (8,3.5);
\draw[dotted] (-1,6) -- (8,6);
\node at (-1,0.5) {\footnotesize (I)};
\node at (-1,1.75) {\footnotesize (II)};
\node at (-1,3) {\footnotesize (III)};
\node at (-1,4.75) {\footnotesize (IV)};
\node at (-1,6.5) {\footnotesize (V)};
\node at (0,8.1) {$t$};
\node at (9.7,0) {$x^+$};
\draw[very thick,blue] (3.75,4.75) -- (1.25,4.75);
\filldraw[blue] (3.75,4.75) circle (2pt);
\filldraw[blue] (1.25,4.75) circle (2pt);
\draw[very thick,blue] (5.75,4.75) -- (8.2,4.75);
\filldraw[blue] (5.75,4.75) circle (2pt);
\draw[very thick,blue] (3,3) -- (2,3);
\filldraw[blue] (3,3) circle (2pt);
\filldraw[blue] (2,3) circle (2pt);
\draw[very thick,blue] (4,3) -- (8.2,3);
\filldraw[blue] (4,3) circle (2pt);
\draw[very thick,blue] (2.75,1.75) -- (0.75,1.75);
\filldraw[blue] (2.75,1.75) circle (2pt);
\filldraw[blue] (0.75,1.75) circle (2pt);
\draw[very thick,blue] (4.25,1.75) -- (7.75,1.75);
\filldraw[blue] (4.25,1.75) circle (2pt);
\filldraw[blue] (7.75,1.75) circle (2pt);
\draw[very thick,blue] (1.5,0.5) -- (0.5,0.5);
\filldraw[blue] (1.5,0.5) circle (2pt);
\filldraw[blue] (0.5,0.5) circle (2pt);
\draw[very thick,blue] (5.5,0.5) -- (6.5,0.5);
\filldraw[blue] (5.5,0.5) circle (2pt);
\filldraw[blue] (6.5,0.5) circle (2pt);
\draw[very thick,blue] (5.5,6.5) -- (0.5,6.5);
\filldraw[blue] (5.5,6.5) circle (2pt);
\filldraw[blue] (0.5,6.5) circle (2pt);
\draw[very thick,blue] (8.2,6.5) -- (7.5,6.5);
\filldraw[blue] (7.5,6.5) circle (2pt);
\node at (1,-0.5) {$a$};
\node at (6,-0.5) {$b$};
\end{tikzpicture}
\caption{\footnotesize For time $t$, the sets of modes $A_R^{(l)}$ bordered in black, the reflected modes $\hat A^{(l)}_R$ in red and $\tilde A_L^{(l)}$ in green. At a given time the entropy of the modes is given by $S_\text{rad}$ for the blue areas that correspond to $A_R^{(l)}\ominus\tilde A_L^{(l)}$.}
\label{fig14}
\end{center}
\end{figure}
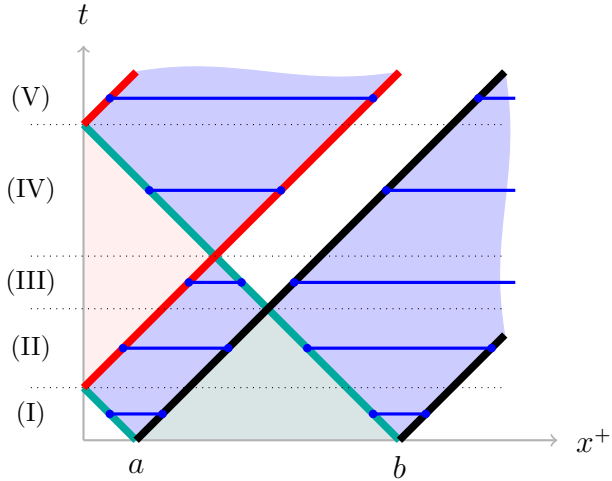

The modes on the left are entangled with modes on the right under the reflection $x^+\to-x^+$, which we define as
\EQ{
x^+\in\tilde A_L^{(l)}=[\max(a-t,0),\max(b-t,0)]\ .
}
The entropy of $A_L\cup A_R$ is then 
\EQ{
S_\text{BCFT}(A_L\cup A_R)=S_\text{rad}\big(\tilde A_L^{(l)}\ominus(A_R^{(l)}\cup\hat A_R^{(l)})\big)+S_\text{rad}\big(\tilde A_R^{(r)}\ominus(A_L^{(r)}\cup\hat A_L^{(r)})\big)\ ,
}
where we have added a similar contribution from the right-moving modes. The entropy of a null interval of radiation is defined in \eqref{radlimit}. The symmetric product $\tilde A_L^{(l)}\ominus(A_R^{(l)}\cup\hat A_R^{(l)})$ is illustrated in figure \ref{fig14} from which the entropy can simply be read off.

\subsection{Black hole and `islands in the stream'}

In this section, we establish a similar geometrical approach to visualize the entanglement structure of the state of the radiation leading to a simple way to calculate the entropy of a saddle in the black hole. This visualization is a generalization of the `islands-in-the-stream' formalism of \cite{Hollowood:2021nlo}.

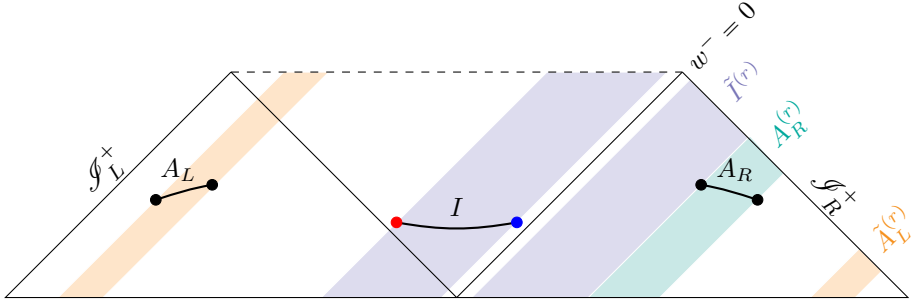
\begin{figure}[ht]
\begin{center}
\begin{tikzpicture}[scale=1]
%\draw[fill=yellow!20,yellow!20] (-6,0) -- (-3,0) -- (-3,3) -- cycle;
%\draw[fill=yellow!20,yellow!20] (6,0) -- (3,0) -- (3,3) -- cycle;
%\draw[white,fill=Plum!10!white] (-3,0) -- (3,0) -- (3,3) -- (-3,3) -- cycle;
%
\draw[white,fill=BurntOrange!20!white] (-5.3,0) -- (-2.3,3) -- (-1.7,3) -- (-4.7,0) -- cycle;
\draw[white,fill=BurntOrange!20!white] (5.3,0) -- (5.65,0.35) -- (5.35,0.65) -- (4.7,0) -- cycle;
\draw[white,fill=Periwinkle!20!white] (-1.8,0) -- (1.2,3) -- (2.8,3) -- (-0.2,0) -- cycle;
\draw[white,fill=Periwinkle!20!white] (1.8,0) -- (3.9,2.1) -- (3.1,2.9) -- (0.2,0) -- cycle;
\draw[white,fill=Emerald!20!white] (1.75,0) -- (3.9,2.15) -- (4.35,1.65) -- (2.7,0) -- cycle;
\draw[-] (-3,3) -- (-6,0) -- (6,0) -- (3,3);
\draw[dashed] (-3,3) -- (3,3);
\draw[-] (0,0) -- (3,3);
\draw[-] (-3,3) -- (0,0);
\filldraw[black] (3.25,1.5) circle (2pt);
\filldraw[black] (-3.25,1.5) circle (2pt);
\filldraw[black] (4,1.3) circle (2pt);
\filldraw[black] (-4,1.3) circle (2pt);
\draw[thick] (-3.25,1.5) to[out=-170,in=18] (-4,1.3);
\draw[thick] (3.25,1.5) to[out=-10,in=162] (4,1.3);
\draw[thick] (-0.8,1) to[out=-10,in=-170] (0.8,1);
\filldraw[blue] (0.8,1) circle (2pt);
\filldraw[red] (-0.8,1) circle (2pt);
\node at (0,1.2) {\footnotesize $I$};
\node at (3.7,1.7) {\footnotesize $A_R$};
\node at (-3.7,1.7) {\footnotesize $A_L$};
\node[rotate=45,Periwinkle] at (3.8,2.9) {\footnotesize $\tilde I^{(r)}$};
\node[rotate=45,Emerald] at (4.4,2.3) {\footnotesize $A_R^{(r)}$};
\node[rotate=45,BurntOrange] at (5.8,0.9) {\footnotesize $\tilde A_L^{(r)}$};
%
%\draw (4.2,4.2) -- (4.3,4.3) -- (7.3,1.3) -- (7.2,1.2);
\node[rotate=-45] at (5,1.4) {$\mathscr I^+_R$};
\node[rotate=45] at (-4.7,1.7) {$\mathscr I^+_L$};
\node[rotate=45] at (3.5,3.5) {\footnotesize $w^-=0$};
\end{tikzpicture}
\caption{\footnotesize The right-moved modes at $\mathscr I^+_R$ that pass through $A_R$ and the purifiers from $A_L\cup I$ that are mapped onto $\mathscr I^+_R$ via the involution about the horizon $w^-=0$ as shown.}
\label{fig15} 
\end{center}
\end{figure}

Take the right-moving modes. Those that pass through $A_R$ are projected onto right null infinity as  $A_R^{(r)}\subset\mathscr I^+_R$ as shown in figure \ref{fig15}. In the large temperature limit, the entropy of these modes, up to the usual UV divergence, is simply the thermodynamic entropy of a gas of massless quanta in the interval defined by the Minkowski null coordinates of the endpoints $x^-\in[t-b,t-a]$ \eqref{radlimit}.
However, the interval $A_L\cup I$ behind the right horizon ($w^-=0$) collects right-moving purifiers of the modes at $\mathscr I^+_R$  and so can cancel the entropy of $A_R^{(r)}$ as well as adding in additional contributions. The purifiers are related by the involution $w^-\to-w^-$, which for an interval of right-moving modes we denote $A^{(r)}\to\tilde A^{(r)}$, and using this we can map the right-moving modes that pass through $A_L\cup I$ onto right null infinity as $\tilde A_L^{(r)}\cup \tilde I^{(r)}\subset\mathscr I^+_R$. Note that the involution preserves the values of the coordinates $x^-$ and so it is simple to find the images of $I$ and $A_L$ on $\mathscr I^+_R$, they are simply the intervals of their $x^-$ coordinates.

There is a subtlety for the islands. If the QES is inside the horizon, i.e.~has coordinate $w^-_\sigma=e^{-2\pi x^-_\sigma/\beta}$, the involution $w^-\to-w^-$ maps it to $-e^{-2\pi x^-_\sigma/\beta}$ and projects onto $\mathscr I^+_R$ as a point with null Minkowski coordinate $x^-_\sigma$. However, if the QES is outside the horizon, i.e.~has coordinate $w^-_\sigma=-e^{-2\pi x^-_\sigma/\beta}$ then the interval $[-w^-_\sigma,w^-_\sigma]\subset I$ that straddles the horizon purifies itself and therefore cancels out as far as the entropy is concerned and so we can effectively replace the QES coordinate by $e^{-2\pi x^-_\sigma/\beta}$ which is inside the horizon.

The net contribution from the right-moving modes corresponds to the thermodynamic entropy of the gas on the interval of the symmetric difference $A_R^{(r)}\ominus(\tilde A_L^{(r)}\cup \tilde I^{(r)})$. The symmetric difference accounts for the purification of modes across the horizon and the resulting nullification of the entropy. This statement is correct up to the fact that we have to compensate for the conformal factors of the QES since these are not those of a thermal state due to the curved geometry of AdS. The final formula for the entropy is
\EQ{
S_I(A_L\cup A_R)=\sum_{\sigma=1_Q,2_Q}S_\text{QES}(\sigma)+S_\text{rad}\big(A_R^{(r)}\ominus(\tilde A_L^{(r)}\cup \tilde I^{(r)})\big)+S_\text{rad}\big(A_L^{(l)}\ominus(\tilde A_R^{(l)}\cup\tilde I^{(l)})\big)\ ,
}
where the contributions from the QES include the compensator for the conformal factor:
\EQ{
S_\text{QES}(\sigma)=\frac{\pi c}{6\beta k}+\frac c{12}\log|w^+_\sigma w^-_\sigma|\ .
}
We have applied the same logic to the left-moving modes projected onto $\mathscr I^+_L$.

\begin{figure}[ht]
\begin{center}
\begin{tikzpicture}[scale=0.7]
\draw[white,fill=red!30!white,opacity=0.2] (0,1) -- (-2.5,3.5) -- (-2.5,6) -- (2.5,6) -- (2.5,3.5) -- (3.5,2.5) -- (2.75,1.75) -- cycle;
\draw[white,fill=Emerald!30!white,opacity=0.2] (-6,0) -- (-1,0) -- (-8,7) to[out=-80,in=100] (-8,2) -- cycle;
\draw[white,fill=black!30!white,opacity=0.2] (-6,0) -- (-1,0) -- (6,7) to[out=-170,in=10] (1,7) -- cycle;
\draw[blue!20,fill=blue!20] (-1,0) -- (0,1) -- (-2.5,3.5) -- (-3.5,2.5) -- cycle;
\draw[blue!20,fill=blue!20] (2.75,1.75) --  (0,1)  -- (2.5,3.5) -- (3.5,2.5) -- cycle;
\draw[blue!20,fill=blue!20] (-6,0) -- (-3.5,2.5) -- (-8,7) to[out=-80,in=100] (-8,2) -- cycle;
\draw[blue!20,fill=blue!20] (-2.5,3.5) -- (-2.5,6) -- (0,6) -- cycle;
\draw[blue!20,fill=blue!20] (2.5,6) -- (2.5,3.5) -- (6,7) to[out=-170,in=10] (1,7) -- (0,6) -- cycle;
\draw[black!30,->,thick] (-8,0) -- (7.5,0);
\draw[black!30,->,thick] (0,0) -- (0,7.5);
\draw[line width=1mm,Emerald] (-1,0) -- (-8,7);
\draw[line width=1mm,Emerald] (-6,0) -- (-8,2);
\draw[line width=1mm] (-1,0) -- (6,7);
\draw[line width=1mm] (-6,0) -- (1,7);
\draw[red,line width=1mm] (0,1) -- (-2.5,3.5) -- (-2.5,6);
\draw[red,line width=1mm] (0,1) -- (2.75,1.75) -- (3.5,2.5) -- (2.5,3.5) -- (2.5,6);
\draw[dotted] (-8,1) -- (7,1);
\draw[dotted] (-8,2.5) -- (7,2.5);
\draw[dotted] (-8,3.5) -- (7,3.5);
\draw[dotted] (-8,6) -- (7,6);
\node at (6.5,0.5) {\footnotesize (I)};
\node at (6.5,1.4) {\footnotesize (IIa)};
\node at (6.5,2.1) {\footnotesize (IIb)};
\node at (6.5,3) {\footnotesize (III)};
\node at (6.5,4.75) {\footnotesize (IV)};
\node at (6.5,6.5) {\footnotesize (V)};
\node at (0,8.1) {$t$};
\node at (8.3,0) {$x^-$};
\node[red] at (0,3.5) {$\tilde I^{(r)}$};
\node at (3.5,7.5) {$A_R^{(r)}$};
\node[Emerald] at (-9.2,4.5) {$\tilde A_L^{(r)}$};
\draw[very thick,blue] (3.75,4.75) -- (2.5,4.75);
\filldraw[blue] (3.75,4.75) circle (2pt);
\filldraw[blue] (2.5,4.75) circle (2pt);
\draw[very thick,blue] (-2.5,4.75) -- (-1.25,4.75);
\filldraw[blue] (-2.5,4.75) circle (2pt);
\filldraw[blue] (-1.25,4.75) circle (2pt);
\draw[very thick,blue] (-5.75,4.75) -- (-8.2,4.75);
\filldraw[blue] (-5.75,4.75) circle (2pt);
\draw[very thick,blue] (3,3) -- (2,3);
\filldraw[blue] (3,3) circle (2pt);
\filldraw[blue] (2,3) circle (2pt);
\draw[very thick,blue] (-3,3) -- (-2,3);
\filldraw[blue] (-3,3) circle (2pt);
\filldraw[blue] (-2,3) circle (2pt);
\draw[very thick,blue] (-4,3) -- (-8.2,3);
\filldraw[blue] (-4,3) circle (2pt);
\draw[very thick,blue] (2.75,1.75) -- (0.75,1.75);
\filldraw[blue] (2.75,1.75) circle (2pt);
\filldraw[blue] (0.75,1.75) circle (2pt);
\draw[very thick,blue] (-2.75,1.75) -- (-0.75,1.75);
\filldraw[blue] (-2.75,1.75) circle (2pt);
\filldraw[blue] (-0.75,1.75) circle (2pt);
\draw[very thick,blue] (-4.25,1.75) -- (-7.75,1.75);
\filldraw[blue] (-4.25,1.75) circle (2pt);
\filldraw[blue] (-7.75,1.75) circle (2pt);
\draw[very thick,blue] (-1.5,0.5) -- (-0.5,0.5);
\filldraw[blue] (-1.5,0.5) circle (2pt);
\filldraw[blue] (-0.5,0.5) circle (2pt);
\draw[very thick,blue] (-5.5,0.5) -- (-6.5,0.5);
\filldraw[blue] (-5.5,0.5) circle (2pt);
\filldraw[blue] (-6.5,0.5) circle (2pt);
\draw[very thick,blue] (5.5,6.5) -- (0.5,6.5);
\filldraw[blue] (5.5,6.5) circle (2pt);
\filldraw[blue] (0.5,6.5) circle (2pt);
\draw[very thick,blue] (-8.2,6.5) -- (-7.5,6.5);
\filldraw[blue] (-7.5,6.5) circle (2pt);
\node at (-1,-0.5) {$-a$};
\node at (-6,-0.5) {$-b$};
\end{tikzpicture}
\caption{\footnotesize The 3 subsets $A_R^{(r)}$ (black), $\tilde A_L^{(r)}$ (green) and $\tilde I^{(r)}$ (red) on $\mathscr I^+_R$ as a function of time. Given that the latter two purify the former, the net contribution to the entropy at any given time is given by $A_R^{(r)}\ominus(\tilde A_L^{(r)}\cup \tilde I^{(r)})$ the blue area (plus the contribution from the QES). Example time slices are shown in each time regime. In regime (III), the blue regions are shrinking due a collision between $\tilde A_L^{(r)}$ and the island in the stream $\tilde I^{(r)}$ as $t$ increases accounting for the dip in the entropy.}
\label{fig16}
\end{center}
\end{figure}
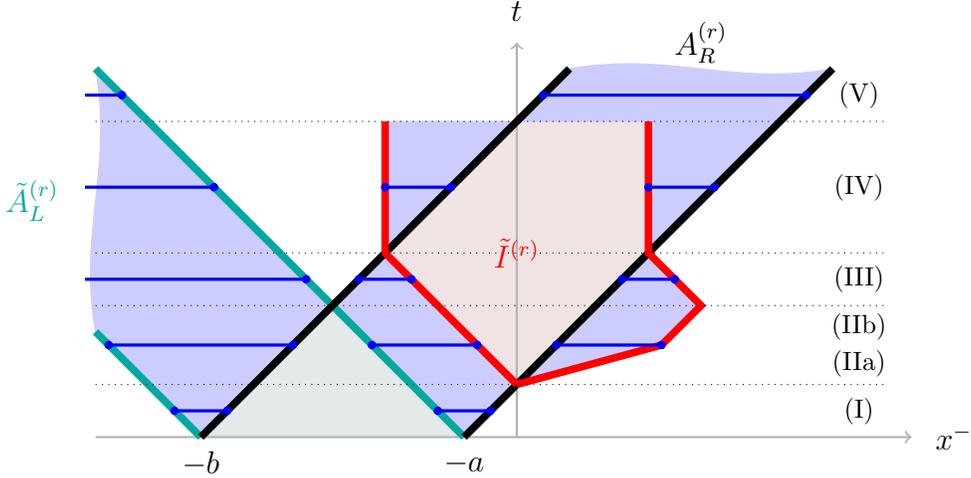

The behaviour of the set $A_R^{(r)}\ominus(\tilde A_L^{(r)}\cup\tilde I^{(r)})$ as a function of time is shown in figure \ref{fig16}. As an example, consider the saddle in regime (III) where the entropy unexpectedly decreases. In that case, the projections onto $\mathscr I^+_R$ in terms of the coordinate $x^-$ are
\EQ{
A_R^{(r)}=[t-b,t-a]\ ,\quad \tilde A_L^{(r)}=[-t-b,-t-a]\ ,\quad \tilde I_R^{(r)}=[-t+a,-t+b]\ .
}
Hence,\footnote{Note that the following intervals are just pairs of the $x^-$ in the ordered list in \eqref{swy2}.}
\EQ{
A_R^{(r)}\ominus(\tilde A_L^{(r)}\cup \tilde I^{(r)})=[-t-b,-t-a]\cup[t-b,-t+a]\cup[t-a,-t+b]
}
and so
\EQ{
S_\text{rad}(A_R^{(r)}\ominus(\tilde A_L^{(r)}\cup \tilde I^{(r)}))=\frac{\pi c}{6\beta}\big((b-a)+(b+a-2t)+(b+a-2t)\big)\ .
}
The left-moving modes give an identical contribution. Note that the second and third intervals decrease in size as $t$ increases which is consistent with the decrease in the entropy. However, there are also the QES contributions to consider
\EQ{
S_\text{QES}(1_Q)=S_\text{QES}(2_Q)=\frac{\pi c}{6\beta k}+\frac{\pi c}{6\beta}(2t-a-b)\ .
}
Adding all the contributions gives the result \eqref{beq}.

\section{Discussion}\label{s6}

In BCFT, the boundary acts as a mirror that reflects incoming modes. For spatial intervals of characteristic length $L$, the thermodynamic limit $L/\beta\gg1$ uncovers behaviour that is captured by competing OPE channels, and correlations on short scales $\sim {\cal O}(\beta)$ are invisible. The only effect of the boundary degrees of freedom is to contribute an additive boundary entropy $s_b=\log g_b$ in the relevant OPE channels. The  free fermion theory is particularly simple given that $\log g_b=0$. Despite the simplicity of this limit, the time evolution of measures of entanglement can be nontrivial.

These generic features are reproduced by  AdS$_2$ black holes in the Hartle-Hawking state with radiation baths, in the limit of large intervals such that the scrambling time scale can  be neglected  $L\gg \beta\log\left(2\pi/k\beta\right)\gg \beta$. If we also take $L\gg 1/k$, we recover the free fermion results with vanishing boundary entropy. In this limit, the black hole is particularly simple and entanglement wedge reconstruction of its interior should be simple. Its entropy is much smaller than the thermodynamic entropy of the intervals and the effects of scrambling being negligible, the black hole behaves like a  mirror. The reflected modes on either side of the thermofield double play the role of the Hawking radiation, and the BCFT images on the left and right play the role of the interior of the black hole.  

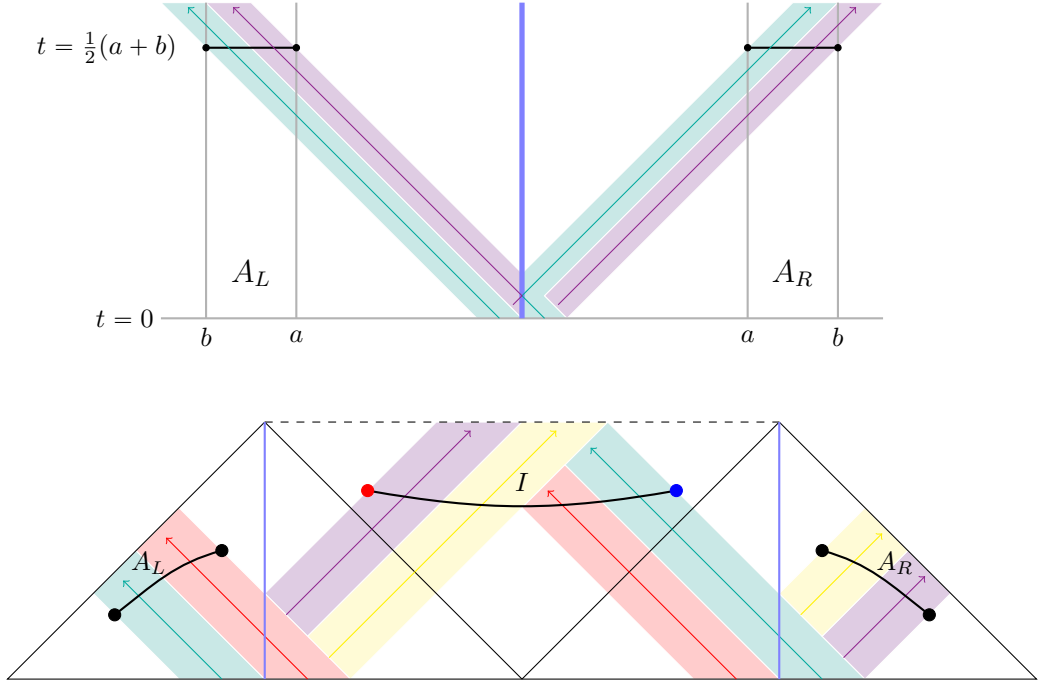
\begin{figure}[ht]
\begin{center}
\begin{tikzpicture}[scale=0.6]
\draw[white,fill=Emerald!20!white] (0,0) -- (-7,7) -- (-8,7) -- (-1,0) -- cycle;
\draw[white,fill=Emerald!20!white] (0,0) -- (1,0) -- (0.5,0.5) -- (7,7) -- (6,7) -- (0,1) -- cycle;
\draw[white,fill=Plum!20!white] (0,0) -- (-7,7) -- (-6,7) -- (0,1) -- cycle;
\draw[white,fill=Plum!20!white] (0.5,0.5) -- (7,7) -- (8,7) -- (1,0) -- cycle;
\draw[black!30,thick] (-8,0) -- (8,0);
\draw[blue!50,line width=0.7mm] (0,0) -- (0,7);
\draw[black!30,thick] (-7,0) -- (-7,7);
\draw[black!30,thick] (-5,0) -- (-5,7);
%\draw[thick] (-1,0) -- (-8,7);
%\draw[thick] (0,1) -- (-6,7);
%
\draw[black!30,thick] (7,0) -- (7,7);
\draw[black!30,thick] (5,0) -- (5,7);
\draw[thick] (-7,6) -- (-5,6);
\draw[thick] (7,6) -- (5,6);
\filldraw[black] (-5,6) circle (2pt);
\filldraw[black] (-7,6) circle (2pt);
\filldraw[black] (5,6) circle (2pt);
\filldraw[black] (7,6) circle (2pt);
\node at (-9.2,6) {\footnotesize $t=\tfrac12(a+b)$};
\node at (-8.8,0) {\footnotesize $t=0$};
\node at (-7,-0.4) {\footnotesize $b$};
\node at (-5,-0.4) {\footnotesize $a$};
\node at (7,-0.4) {\footnotesize $b$};
\node at (5,-0.4) {\footnotesize $a$};
\node at (-6,1) {$A_L$};
\node at (6,1) {$A_R$};
\draw[Emerald,->] (0.5,0) -- (0,0.5) -- (6.4,6.9);
\draw[Plum,->] (0.8,0.3) -- (7.4,6.9);
\draw[Emerald,->] (-0.5,0) -- (-7.4,6.9);
\draw[Plum,->] (-0.2,0.3) -- (0,0.5) -- (-6.4,6.9);
\begin{scope}[yshift=-8cm,scale=1.9]
\draw[white,fill=Emerald!20!white] (4,0) -- (3,0) -- (0.5,2.5) -- (1,3) -- cycle;
\draw[white,fill=Emerald!20!white] (-3,0) -- (-4,0) -- (-5,1) -- (-4.5,1.5) -- cycle;
\draw[white,fill=Plum!20!white] (3.5,0.5) -- (4,0) -- (5,1) -- (4.5,1.5) -- cycle;
\draw[white,fill=Plum!20!white] (-3,1) -- (-2.5,0.5) -- (0,3) -- (-1,3) -- cycle;
\draw[white,fill=yellow!20!white] (3.5,0.5) -- (3,1) -- (4,2) -- (4.5,1.5) -- cycle;
\draw[white,fill=yellow!20!white] (-2,0) -- (-2.5,0.5) -- (0,3) -- (1,3) -- cycle;
\draw[white,fill=red!20!white] (-3,0) -- (-2,0) -- (-4,2) -- (-4.5,1.5) -- cycle;
\draw[white,fill=red!20!white] (2,0) -- (3,0) -- (0.5,2.5) -- (0,2) -- cycle;
\draw[-] (-3,3) -- (-6,0) -- (6,0) -- (3,3);
\draw[dashed] (-3,3) -- (3,3);
\draw[-] (0,0) -- (3,3);
\draw[-] (-3,3) -- (0,0);
\draw[Emerald,->] (3.5,0) -- (0.8,2.7);
\draw[Plum,->] (3.75,0.25) -- (4.7,1.2);
\draw[Emerald,->] (-3.5,0) -- (-4.65,1.15);
\draw[Plum,->] (-2.75,0.75) -- (-0.6,2.9);
\draw[red,->] (2.5,0) -- (0.3,2.2);
\draw[yellow,->] (3.25,0.75) -- (4.2,1.7);
\draw[red,->] (-2.5,0) -- (-4.15,1.65);
\draw[yellow,->] (-2.25,0.25) -- (0.4,2.9);
\filldraw[black] (3.5,1.5) circle (2pt);
\filldraw[black] (-3.5,1.5) circle (2pt);
\filldraw[black] (4.75,0.75) circle (2pt);
\filldraw[black] (-4.75,0.75) circle (2pt);
\draw[thick] (-3.5,1.5) to[out=-165,in=38] (-4.75,0.75);
\draw[thick] (3.5,1.5) to[out=-15,in=142] (4.75,0.75);
\draw[thick] (-1.8,2.2) to[out=-10,in=-170] (1.8,2.2);
\filldraw[blue] (1.8,2.2) circle (2pt);
\filldraw[red] (-1.8,2.2) circle (2pt);
\node at (0,2.3) {\footnotesize $I$};
\node at (4.35,1.35) {\footnotesize $A_R$};
\node at (-4.35,1.35) {\footnotesize $A_L$};
\draw[blue!50,thick] (3,0) -- (3,3);
\draw[blue!50,thick] (-3,0) -- (-3,3);
\end{scope}
\end{tikzpicture}
\caption{\footnotesize The explanation for the dip in BCFT (top) and black hole (bottom). The coloured outgoing modes are maximally entangled between left and right and so these modes do not contribute to $S(A_L\cup A_R)$. It follows that the entropy is only one half of the thermodynamic entropy of the two intervals. Note that in the BCFT case, the green (purple) modes on the right (left) side start as left- (right-)moving modes at $t=0$ that reflect off the boundary whereas in the black hole there are four distinct sets of modes.}
\label{fig17}
\end{center}
\end{figure}

It is remarkable to see the nontrivial dynamical entanglement  structure being reproduced by plunging QES and the qualitative features continuing to apply away from the strict BCFT limits. In this context, the dip, readily apparent in figures \ref{fig6} and \ref{fig7}, in entanglement entropy of the two disjoint intervals $A_L\cup A_R$ deserves special mention (equivalently a peak in the mutual information) particularly in the case with small intervals. In this case information equilibrium appears at early times after a short period of linear growth of the entropy, but a sudden brief dip appears at a much later time before equilibrium sets in again. What is the physical reason for this dip? The explanation is that precisely at this time the out-going modes, i.e.~left-moving through $A_L$ and right-moving through $A_R$, are completely purified leaving only the contribution from the in-going modes, i.e.~half the number of modes. A picture of the relevant modes for both the BCFT and black hole are shown in figure \ref{fig17}. In the BCFT, the purification can be seen immediately from the geometry. In the black hole case, the QES are positioned in precisely the right place (to leading order) to capture the relevant modes to purify the state.

A noteworthy outcome of the precise comparison between BCFT and gravity is the identification of different disconnected  BCFT channels (corresponding to singularities in the Lorentzian correlator) with  {\em bulk} channels in JT-gravity in the presence of  QES inside or outside the horizon. From the examples  studied in this paper (and in the limit of large intervals and small scrambling times), we infer that the QES serve to reproduce contractions involving BCFT image points in either copy  of the TFD state.  It would be interesting to perform a more exhaustive study of all possible singularity channels in the BCFT picture (for our two interval problem there are 24 such channels) and the precise characterization of their associated QES. It is interesting to note that the purification of modes and consequent dip in entanglement entropy always involves QES behind the horizon. This is similar to the nonequilibrium situation with  evaporating  black holes \cite{Hollowood:2020cou} where the QES remains inside the horizon while the entropy relaxes after Page time and only pops out of the horizon at parametrically late times when the system is approaching equilibrium.

An interesting aspect  of our analysis of the island saddles in gravity is that the QES evolve smoothly across the  different regimes which we have identified, including when they plunge into the horizon, the boundaries between the regimes becoming sharp only in the limit of large intervals. In this sense, the JT gravity realization of island saddles appears different to  large-$c$ holographic BCFTs wherein the  distinct OPE channels are captured by distinct RT surfaces or saddles.
Also important here is that the JT gravity analysis in this paper captures the physics of CFTs with a free quasiparticle description which is distinct from large-$c$ holographic BCFTs that are maximally scrambling and are not expected to exhibit entanglement dips. However, the JT gravity plus free fermion setup also allows the exploration of the strong scrambling regime, where it is the black hole that does the scrambling instead of the CFT bath. In particular, it would be interesting to relax the limit of high temperatures and small scrambling times to understand both the deviations from the BCFT picture and the effects of black hole scrambling on the entanglement evolution.

Finally, it is worth recalling that we have found in our analysis of both the semi-infinite and finite intervals, a time scale associated to  QES exiting the horizon. This timescale $(t_{\rm exit}=2a)$ is much larger than the entanglement equilibrium scale, and has little or no effect on the entropy evolution itself. It would be interesting to learn what physical significance, if any, can be attached to exiting QES and the associated timescale.
%This smooth interpolation between different BCFT-like channels should continue to apply away from the simplifying limit including when the holographic dual of the gravity system may be described by the SYK model coupled to free fermions (see e.g.~\cite{Chen:2020wiq}). 

\vspace{0.5cm}
\begin{center}{\it Acknowledgments}\end{center}
\vspace{0.2cm}
TJH, AL and SPK acknowledge support from STFC grant ST/T000813/1. NT acknowledges the support of an STFC Studentship.

\end{document}